\documentclass[]{bytedance_seed}
\usepackage[utf8]{inputenc}
\usepackage{setspace}
\usepackage{graphicx}
\graphicspath{ {./figures/} }
\usepackage{amsmath}
\usepackage{lineno}
\usepackage{hyperref}
\usepackage{threeparttable}
\usepackage{booktabs}
\usepackage{natbib}
\usepackage{rotating}

\usepackage{xcolor}
\newcommand{\FIXME}[1]{\textcolor{red}{FIXME: #1}}

\title{TDDFT Gradients and Nonadiabatic Couplings with Minimal Auxiliary Basis Set Approximation for Fewest-Switches Surface Hopping Dynamics}

\author[1,2,\dag]{Cheng Fan}
\author[1,\dag]{Zhichen Pu}
\author[3]{Zehao Zhou}
\author[1]{Yuanheng Wang}
\author[2,*]{Yi Qin Gao}
\author[1,*]{Qiming Sun}

\affiliation[1]{ByteDance Seed}
\affiliation[2]{New Cornerstone Science Laboratory, College of Chemistry and Molecular Engineering, Peking University, Beijing 100871, China}
\affiliation[3]{Zhongguancun Academy}
\contribution[\dag]{These authors contributed equally to this work}
\contribution[*]{Corresponding authors}

\correspondence{Qiming Sun at \email{qiming.sun@bytedance.com}, Yiqin Gao at \email{gaoyq@pku.edu.cn}}
\date{\today}

\abstract{
The electronic structure calculations remain a major bottleneck in ab initio nonadiabatic molecular dynamics.
We develop an efficient TDDFT-based FSSH implementation in the GPU4PySCF package for medium-sized molecular systems.
Our approach combines density fitting, TDDFT with minimal auxiliary basis sets (TDDFT-ris), and an approximate Z-vector solver to reduce the computational cost of TDDFT excited states and derivative coupling calculations.
These approximations introduce negligible errors in realistic FSSH workloads while maintaining high computational efficiency.
Benchmark results show that, for 73-atom systems with a triple-$\zeta$ basis set, individual electronic structure calculations are completed within one minute on a single NVIDIA A100 GPU.
}

\begin{document}
\maketitle


\section{Introduction}
The study of excited-state dynamics is fundamental to understanding a wide range of photochemistry and photophysics processes. These processes require accurate modeling of non-adiabatic dynamics. Among various computational approaches, the trajectory surface-hopping (TSH) method stands out as a prominent quantum-classical hybrid technique to model the non-adiabatic dynamics of molecular systems. The FSSH (Fewest-Switches Surface-Hopping) method, introduced by Tully in 1990 \cite{Tully1990}, is the most widely used and efficient variant of TSH.
This method treats the nuclei classically while modeling electronic transitions quantum mechanically, avoiding the need for a full quantum treatment of the nuclear motion.
Due to its simplicity and effectiveness, it has been widely adopted to study excited-state molecular processes.\cite{tully2012perspective,furche2011,Barbatti2011,Parker2017,Tapavicza2013,Crespo-Otero2018,Song2020,Peters2020,Nelson2020,Wang2020,pyxaid,hefeinamd,newtonx,SHARC,pyrai2md,jadenamd,Myers2024,Peng2022,Qiu2023,Vogt2025}.

Performing the classical trajectory in FSSH is efficient.
The major computational challenge in FSSH simulations is the electronic structure calculation, particularly in the computation of derivative couplings.
Wave-function based methods for computing derivative couplings are often computationally expensive or even prohibitive.
Density functional theory (DFT) offers a computationally efficient alternative, providing a good balance between accuracy and cost.
Time-dependent DFT (TDDFT)\cite{Martin2005,Casida2009}, in particular, has emerged as a promising method to treat excited-state dynamics\cite{tapavicza2007trajectory} with reasonable computational cost, enabling simulations of molecules containing hundreds of atoms.\cite{furche2011,Peters2019,Peters2020,Myers2024}

Recently, we developed analytic gradients and derivative couplings for TDDFT-ris, a minimal auxiliary basis approximation to TDDFT\cite{Pu2026}.
This approximation enables efficient excited-state calculations, which often converges faster than the ground-state SCF computations\cite{Zhou2023,zhou2024converging,zhou2026}.
For nuclear gradients and derivative couplings of excited states\cite{Furche2002,rappoport2005analytical,scalmani2006geometries,herbert2016beyond,zhang2014analytic,zhang2015analytic,send2010first,subotnik2015requisite,zhang2021nonadiabatic,ou2015first,li2014first,li2014first2}, the TDDFT-ris approximation provides reliable results near equilibrium geometries and achieves roughly a two-fold speedup over canonical TDDFT\cite{Pu2026}.
However, larger deviations are observed in more challenging scenarios, such as conical intersection searches, particularly for minimum energy crossing points (MECPs), where the electronic states become quasi-degenerate.
Since FSSH dynamics can propagate to the near-degenerate regions, these discrepancies may become more pronounced, potentially affecting the accuracy of nonadiabatic simulations.
To assess the impact of these deviations on practical simulations, in this work, we evaluate the performance and reliability of TDDFT-ris approximation in FSSH simulations for larger and more complex systems.

Many NAMD packages\cite{pyxaid,hefeinamd,SHARC,newtonx,pyrai2md,jadenamd,Myers2024} offers an interface to invoke external electronic structure programs to obtain excited-state energies, forces, and derivative couplings.
This execution model makes it difficult to maintain and track the electronic structure state across time steps.
Such state information is important for determining the phase of
derivative couplings\cite{Pittner2009, Plasser2016, Ryabinkin2015, Lee2019, Lee2021,Chen2022} and can also improve the efficiency of electronic structure calculations\cite{Peters2019}.
Moreover, when using GPU programs for electronic structure calculations, the inter-process communication can introduce additional overhead.
Modern GPU libraries often rely on just-in-time (JIT) compilation to achieve optimal runtime performance\cite{nvidia_nvcc,nvidia_cutensor,nvidia_cutile,wu2025designingQCalgorithms} but this can introduce noticeable initialization costs.

To address the interface issue, we implement the FSSH algorithm directly within the GPU4PySCF package\cite{Wu2024,li2025introducing}.
This native implementation allows computational intermediates to be reused in a straightforward manner and eliminates the inter-process communication overhead.
It also provides a flexible platform for integrating various excited-state methods developed in PySCF\cite{Sun2020} and GPU4PySCF.
The details of our implementation will be described in this manuscript.

This manuscript is organized as follows: In Section \ref{sec:theory}, we provide a summary of the FSSH method and the TDDFT formulas. Section \ref{sec:implementation} describes the implementation of the native FSSH algorithm as well as the computation considerations of derivative couplings within GPU4PySCF.
Section \ref{sec:performance} presents performance benchmarks, comparing the computational efficiency and accuracy of gradients and derivative couplings computed using canonical TDDFT and various TDDFT-ris approximations.
In Section \ref{sec:applications}, we illustrate three applications involving internal conversion between excited states using our FSSH implementation and assess the accuracy of TDDFT-ris relative to canonical TDDFT methods.

\section{Methods}
\label{sec:theory}
\subsection{Fewest Switches Surface Hopping Algorithm}
\label{sec:fssh}
In non-adiabatic molecular dynamics, the total electronic wavefunction $\Psi(\mathbf{r}_e, t; \mathbf{R}(t))$ is expanded in the adiabatic basis $\{\phi_j(\mathbf{r}_e; \mathbf{R}(t))\}$:
\begin{align}
\Psi(\mathbf{r}_e, t; \mathbf{R}(t)) = \sum_j c_j(t) \, \phi_j(\mathbf{r}_e; \mathbf{R}(t)),
\end{align}
where $c_j(t)$ is the time-dependent expansion coefficient for state $j$, which is determined by 
\begin{align}
i\hbar \dot{c}_k(t) = \sum_j \left[ E_{kj}(\mathbf{R}) - i\hbar \dot{\mathbf{R}}(t) \cdot \mathbf{d}_{kj}(\mathbf{R}) \right] c_j(t).
\end{align}
$E_{kj}(\mathbf{R}) = \langle \phi_k(\mathbf{r}_e; \mathbf{R}) | \hat{H}_e | \phi_j(\mathbf{r}_e; \mathbf{R}) \rangle$ is the electronic Hamiltonian matrix element. In the adiabaitc representation, $E_{kj}$ vanishes when $k \neq j$. $\mathbf{d}_{kj} = \langle \phi_k(\mathbf{r}_e; \mathbf{R}) | \frac{d}{d\mathbf{R}} \phi_j(\mathbf{r}_e; \mathbf{R}) \rangle$ is the derivative coupling between states $k$ and $j$.

In the FSSH framework\cite{Tully1990, Tully1998}, the nuclei are treated as classical particles evolving along Newtonian trajectories on a single adiabatic potential energy surface (PES) $E_k(\mathbf{R})$
\begin{align}
M_I \ddot{\mathbf{R}}_I = -\nabla_{\mathbf{R}_I} E_{k}(\mathbf{R}) = \mathbf{F}_{I,k}(\mathbf{R}),
\end{align}
where $M_I$ and $\mathbf{R}_I$ are the mass and position of the $I$-th nucleus, and $\mathbf{F}_{I,k}(\mathbf{R})$ is the corresponding nuclear force on nucleus $I$ for state $k$.

The probability of an instantaneous switch from the active state $k$ to another state $l$ within a time step $\Delta t$ is calculated as
\begin{align}
P_{k \rightarrow l}(t) = \max\left(0, \frac{2\text{Re}\left[\dot{\mathbf{R}}(t) \cdot \mathbf{d}_{kl}(\mathbf{R}) c_k^*(t) c_l(t)\right] - \frac{2}{\hbar}\text{Im}\left[E_{lk}(\mathbf{R}) c_k^*(t) c_l(t)\right]}{|c_k(t)|^2} \Delta t\right).
\end{align}
Following the ``fewest switches'' criterion, a hop to state $l$ occurs if a randomly generated number $\xi \in [0, 1]$ satisfies:
\begin{align}
\sum_{j=1}^{l-1} P_{k\rightarrow j}(t) < \xi < \sum_{j=1}^l P_{k\rightarrow j}(t).
\end{align}

Energies $E_j$, gradients $\mathbf{F}_j$, and nonadiabatic couplings $\mathbf{d}_{jk}$ are the three key physical quantities obtained from electronic structure calculations.
They can be computed on-the-fly using time-dependent density functional theory (TDDFT).

\subsection{TDDFT Excited-State Properties}
\label{sec:tddft}
In linear-response TDDFT \cite{casida1995time, Martin2005,Casida2009}, the many-body electronic excited-state problem is formulated as a single-particle transition problem by considering the linear response of the ground-state electron density to a time-dependent perturbation.
Excitation energies $\omega$ are obtained by solving Casida's non-Hermitian eigenvalue equation:
\begin{equation}
\begin{pmatrix}
\mathbf{A} & \mathbf{B} \\
\mathbf{B} & \mathbf{A}
\end{pmatrix}
\begin{pmatrix}
\mathbf{X} \\
\mathbf{Y}
\end{pmatrix}
= \omega
\begin{pmatrix}
\mathbf{1} & \mathbf{0} \\
\mathbf{0} & -\mathbf{1}
\end{pmatrix}
\begin{pmatrix}
\mathbf{X} \\
\mathbf{Y}
\end{pmatrix},
\label{equ:casida}
\end{equation}
where $\mathbf{X}$ and $\mathbf{Y}$ denote the excitation and de-excitation transition amplitude vectors, respectively. The matrix elements of $\mathbf{A}$ and $\mathbf{B}$ are defined as:
\begin{align}
    A_{ai, bj} =& (\varepsilon_a - \varepsilon_i) \delta_{ab} \delta_{ij} + K_{ai, bj}, \label{equ:A_matrix} \\
    B_{ai, bj} =& K_{ai, jb}, \label{equ:B_matrix} \\
    K_{pq,rs} =& g_{pq,sr} + (f_\text{xc})_{pq,sr}, \label{equ:K_matrix} \\
    g_{pq,sr} =& (pq|sr) - c_x(pr|sq) \label{equ:g_integral}, \\
    (f_\text{xc})_{pq,sr} = & \frac{\partial^2 E_{\text{xc}}}{\partial D_{qp} \partial D_{rs}}, \label{equ:g_xc} \\
    (pq|sr) = & \iint \mathrm{d} \boldsymbol{r}_1 \mathrm{d} \boldsymbol{r}_2  \frac{p(\boldsymbol{r}_1) q(\boldsymbol{r}_1) s(\boldsymbol{r}_2) r(\boldsymbol{r}_2)}{|\boldsymbol{r}_1 - \boldsymbol{r}_2|}  , \label{equ:pqrs}
\end{align}
Here, $i, j$ represent occupied orbitals, $a, b$ denote unoccupied orbitals, and $p, q$ stand for general molecular orbitals.
$f_{\text{xc}}$ is the exchange-correlation kernel.
Chemist's notation $(pq|rs)$ is used for the two-electron repulsion integral.
Here, we follow the quantum chemistry indexing convention, using lower-case indices to label molecular orbitals and upper-case indices to label electronic excited states.
These notations should be distinguished from electronic state indices (also lower-case) and the nuclear degrees of freedom (upper-case) employed in the time-dependent Schr\"odinger equation in Section~\ref{sec:fssh}.

In practical calculations, the Tamm-Dancoff approximation (TDA) is often adopted.
The TDA method neglects the contribution of the de-excitation matrix $\mathbf{B}$.
Casida's equation~\eqref{equ:casida} is then simplified to a Hermitian eigenvalue problem:
\[
\mathbf{A}\mathbf{X} = \omega\mathbf{X}
\]

The gradients of the 
$I$-th electronic state with respect to nuclear coordinates is
\[
g_{I}^\xi = \frac{\partial}{\partial \xi} E_I, \quad \xi \in x = \{\boldsymbol{R}_1, \boldsymbol{R}_2, \ldots, \boldsymbol{R}_N\},
\]
where $ E_I $ denotes the energy of the $I$-th electronic state.
In addition, the derivative coupling between electronic states 
$I$ and $J$ is defined as
\[
g_{IJ}^\xi = \langle \Psi_I | \nabla_\xi \Psi_J \rangle.
\]
The expressions for excited-state gradients as well as derivative couplings between ground and excited states, or a pair of excited states, can be written in a unified form
\begin{align}
  g_I^{\xi} &= \langle \boldsymbol{h}^{(\xi)}; \boldsymbol{P}_I \rangle + \langle \boldsymbol{g}^{(\xi)}; \{ \boldsymbol{D}, \boldsymbol{P}_I \} + \boldsymbol{\varGamma}_{II} \rangle
  + \langle \boldsymbol{v}_{\text{xc}}^{(\xi)}; \boldsymbol{P_I} \rangle + \langle \boldsymbol{f}_{\text{xc}}^{(\xi)}; \{ \boldsymbol{R}_I, \boldsymbol{R}_I \} \rangle 
  - \langle \boldsymbol{S}^{(\xi)}; \boldsymbol{W} \rangle. \label{equ:gI_xi} \\
  g_{0I}^{\xi} &= \langle \boldsymbol{h}^{(\xi)}; \boldsymbol{Z} \rangle + \langle \boldsymbol{g}^{(\xi)}; \{ \boldsymbol{D}, \boldsymbol{Z} \} \rangle 
  + \langle \boldsymbol{v}_{\text{xc}}^{(\xi)}; \boldsymbol{Z} \rangle - \langle \boldsymbol{S}^{(\xi)}; \boldsymbol{W} \rangle.\\
  g_{IJ}^{\xi} & = \frac{\tilde{L}_{IJ}^{(\xi)}}{E_J-E_I} , \\
  \tilde{L}_{IJ}^{(\xi)}  &= \langle \boldsymbol{h}^{(\xi)}; \boldsymbol{P}_{IJ} \rangle + \langle \boldsymbol{g}^{(\xi)}; \{ \boldsymbol{D}, \boldsymbol{P}_{IJ} \} + \boldsymbol{\varGamma}_{IJ} \rangle
  + \langle \boldsymbol{v}_{\text{xc}}^{(\xi)}; \boldsymbol{P}_{IJ} \rangle + \langle \boldsymbol{f}_{\text{xc}}^{(\xi)}; \{ \boldsymbol{R}_I, \boldsymbol{R}_J \} \rangle
  - \langle \boldsymbol{S}^{(\xi)}; \boldsymbol{W} \rangle.
  \label{equ:gIJ_xi}
\end{align}
Here,
$\boldsymbol{h}$ denotes the one-electron Hamiltonian,
$\boldsymbol{g}$ represents the two-electron integral tensor, and
$\boldsymbol{S}$ is the atomic orbital overlap matrix.
$\boldsymbol{D}$ is the ground state density matrix.
$\boldsymbol{P}_I$, $\boldsymbol{P}_{IJ}$, $\boldsymbol{R}_{I}$, and $\boldsymbol{W}$ are
intermediates that have a structure similar to the density matrix.
$\boldsymbol{Z}$ is the Lagrange multiplier obtained by solving the Z-vector equation\cite{Handy1984}.
A direct evaluation of $g_I^\xi$ and $g_{IJ}^\xi$
involves derivatives of molecular orbital coefficients with respect to nuclear coordinates, which formally requires solving the coupled-perturbed Kohn–Sham (CPKS) equations for each of the
$3N$ nuclear degrees of freedom.
By using the Lagrangian formalism\cite{Furche2002},
the explicit solution of the CPKS equations for nuclear
perturbations is replaced by a single Z-vector equation
\begin{align}
    (\boldsymbol{A} + \boldsymbol{B}) \boldsymbol{Z} = g^{(\mathrm{VO})} - g^{(\mathrm{OV})}.
    \label{equ:Z_vector}
\end{align}
The right-hand side of the Z-vector equation varies according to the required property.
A detailed definition of the notations in Eqs. \eqref{equ:gI_xi}--\eqref{equ:Z_vector} has been provided in our previous work\cite{Pu2026}.

\subsection{Minimal Auxiliary Basis Set Approximation}
The primary computational bottleneck in solving the TDDFT Casida equation~\eqref{equ:casida} lies in the construction of the coupling matrix $\boldsymbol{K}$ \eqref{equ:K_matrix}.
This step requires the evaluation of extensive two-electron integrals, as well as the numerical integration of the XC kernel.
To reduce this computational cost, the TDDFT-ris approach adopts the resolution of identity (RI) approximation, utilizing a compact auxiliary basis set to represent the two-electron integrals within the coupling matrix.
Furthermore, this approach neglects the contribution arising from the kernel of the pure XC functional.
These approximations can be expressed as
\begin{align}
    K_{pq,rs} \approx & K_{pq,rs}^\text{\text{ris}}  =  g^\text{\text{ris}}_{pq,sr}, \label{equ:ris} \\
    g^\text{\text{ris}}_{pq,sr} = & \sum_{AB}(pq|A)(\boldsymbol{M}^{-1})_{AB}(B|sr)
    - c_x \sum_{AB}(pr|A)(\boldsymbol{M}^{-1})_{AB}(B|sq), \label{equ:g_modified} \\
    M_{AB} &= (A|B),
\end{align}
where the indices $A$ and $B$ refer to the auxiliary basis functions, and $(A|B)$ and $(pq|A)$ denote the two-center and three-center two-electron repulsion integrals, respectively.

Within the TDDFT-ris approximation, the expressions for excited-state gradients and derivative couplings can be simplified to
\begin{align}
  g_I^\xi
  = & \langle \boldsymbol{h}^{(\xi)}; \boldsymbol{P}_I \rangle 
  + \langle \boldsymbol{g}^{(\xi)}; \{ \boldsymbol{D}, \boldsymbol{P}_I \} \rangle 
  + \langle \boldsymbol{g}^{\text{ris},(\xi)}; \boldsymbol{\varGamma}_{II} \rangle
  + \langle \boldsymbol{v}_{\text{xc}}^{(\xi)}; \boldsymbol{P}_I \rangle 
  - \langle \boldsymbol{S}^{(\xi)}; \boldsymbol{W} \rangle \label{equ:gI_xi_ris}, \\
  g_{IJ}^{\xi} 
  = & \frac{1}{E_J - E_I}  \left( \langle \boldsymbol{h}^{(\xi)}; \boldsymbol{P}_{IJ} \rangle 
  + \langle \boldsymbol{g}^{(\xi)}; \{ \boldsymbol{D}, \boldsymbol{P}_{IJ} \} \rangle
  + \langle \boldsymbol{g}^{\text{ris},(\xi)}; \boldsymbol{\varGamma}_{IJ} \rangle
  + \langle \boldsymbol{v}_{\text{xc}}^{(\xi)}; \boldsymbol{P}_{IJ} \rangle  - \langle \boldsymbol{S}^{(\xi)}; \boldsymbol{W}\right).
  \label{equ:gIJ_xi_ris}
\end{align}
It should be noted that the \emph{ris} approximation is not applied to every term that involves two-electron integrals or XC contributions.
In particular, the two-electron terms in the orbital Hessian matrix of the Z-vector equations are derived from the ground-state orbital response.
They remain identical to those in canonical TDDFT.

\subsection{Approximated Gradients and Derivative Couplings}
The strict treatment of two-electron contributions in the Z-vector equations \eqref{equ:Z_vector} dominates the computational cost in the evaluation of derivative couplings.
To reduce this cost, we follow the spirit of the TDDFT-ris approximation and further approximate the two-electron terms in the orbital Hessian of the Z-vector equation using a minimal auxiliary basis set \eqref{equ:ris}.

The effectiveness of this approximation can be understood by examining the structure of the orbital Hessian matrix.
The diagonal elements of the orbital Hessian matrix, given by the orbital energy differences between orbital pairs, provide the dominant contribution, while the two-electron terms are generally smaller corrections.
Therefore, approximating these two-electron contributions using a minimal auxiliary basis introduces only higher-order errors in the TDDFT Casida equation \eqref{equ:casida} and the Z-vector equations \eqref{equ:Z_vector}.

It should be noted that this approximation can break the
consistency between nuclear gradients and total energies, i.e., the computed gradients no longer match the numerical derivatives of the energy with respect to nuclear coordinates.
As a result, this approximation is not suitable for geometry optimization workflows, where strict energy–gradient consistency is required.
In contrast, for nonadiabatic molecular dynamics, one would expect that the impact of approximate derivative couplings on hopping probabilities is moderate.

The minimal auxiliary basis RI approximation are consider at
different levels: to the TDDFT Casida equation \eqref{equ:casida} (yielding the TDDFT-ris method), to the
Z-vector equation \eqref{equ:Z_vector} (referred to here as the ris-Z-vector approximation), or to both.
Their accuracy and computational performance will be assessed quantitatively in Section~\ref{sec:performance}.
It is inappropriate to apply such minimal auxiliary basis approximations to other two-electron terms.
Doing so would lead to significantly larger errors in both gradients and derivative couplings.
Therefore, we do not further consider such approximations.

\section{Implementations}
\label{sec:implementation}
\subsection{Gradients and Derivative Couplings}
\label{sec:nacv:implementation}
Nonadiabatic molecular dynamics typically requires the evaluation of derivative couplings between multiple excited states and nuclear gradients of excited states.
In Ehrenfest dynamics, this involves pairwise derivative couplings among multiple states
together with their gradients, whereas FSSH requires derivative couplings between the active state and all others, along with the gradient of the active state.

In TDDFT, derivative couplings and gradients for different states share various intermediates that involve the evaluation of two-electron integrals.
Evaluating them separately leads to redundant computation of these integrals.
Due to the similarity between the working equations for gradients and derivative couplings [Eqs.~\eqref{equ:gI_xi} -- \eqref{equ:gIJ_xi}],
we combine the computation of derivative couplings and excited-state gradients for multiple states into a single routine to reduce this overhead.

In this routine, the most demanding terms are $\langle \boldsymbol{g}; \{\boldsymbol{D}, \boldsymbol{Z}\}\rangle$, $\langle \boldsymbol{g}; \{\boldsymbol{D}, \{\boldsymbol{P}_{IJ}\} + \boldsymbol{\Gamma}_{IJ} \rangle$, and $\langle \boldsymbol{g}; \{\boldsymbol{D}, \{\boldsymbol{P}_I\} + \boldsymbol{\Gamma}_{II}\rangle$.
These terms can ultimately be expressed as contractions between four-index two-electron integrals and density-matrix pairs\cite{Pu2026}.
Our implementation collects all required density-matrix pairs and processes them in a single contraction routine, in which the two-electron integrals are evaluated only once and applied to all density matrices.

When the four-center two-electron integrals are evaluated analytically, the combined evaluation is straightforward.
For each integral generated, we loop over all density-matrix pairs and perform on-the-fly contraction within the GPU kernel.
The situation becomes more challenging when integrals are evaluated using the density fitting approximation.
In this case, the exchange-type contributions require tensor contractions involving five tensors
\begin{gather}
 \langle \boldsymbol{g}; \{\boldsymbol{D},\boldsymbol{D}'\}\rangle
 = \sum_{ABpqrs} (pq|A) (\boldsymbol{M}^{-1})_{AB} (B|sr) D_{qs} D'_{pr}.
\end{gather}
A straightforward implementation would require constructing intermediate three-index tensors of size
$N_\text{Aux} N_\text{AO}^2$, which quickly exceeds GPU memory limits when multiple derivative couplings and gradients are evaluated simultaneously.

To address this memory issue, we apply singular value decomposition (SVD) to the density matrices, retaining only vectors associated with singular values larger than $10^{-8}$.
Each density matrix is thus represented as a product of two tall matrices of size $N_\text{AO} \times N_\text{SVD}$,
\begin{equation}
    D_{pr} = \sum_i C_{pi} C_{ri},
\end{equation}
where $N_\text{SVD}$ is typically only 1/5 to 1/3 of $N_\text{AO}$.
By rearranging the contracting order,
\begin{gather}
    T_{Aij} = \sum_{pq} C_{pi} C_{qj} (pq|A), \\
    \langle \boldsymbol{g}; \{\boldsymbol{D},\boldsymbol{D}'\}\rangle
    = \sum_{ABij} T_{Aij} (\boldsymbol{M}^{-1})_{AB} T_{Bji},
\end{gather}
the size of the intermediates can be reduced to $N_\text{Aux} N_\text{SVD}^2$.

Despite the use of SVD to compress the intermediates, GPU memory remains a limiting factor.
When the available GPU memory is insufficient to store all intermediates simultaneously, the density-matrix pairs are partitioned into batches and processed separately.
As a result, the two-electron integrals must be reevaluated for each batch.
Due to this limitation, when the density fitting approach is applied to large molecular systems, the efficiency of the combined evaluation of gradients and derivative couplings could be reduced.

In the TDDFT-ris approximation, as shown in Eqs.~\eqref{equ:gI_xi_ris} and \eqref{equ:gIJ_xi_ris}, the contributions that involve the ground-state density matrix are evaluated explicitly, while the remaining terms are treated using the minimal RI approximation.
The number of required batches for these ground-state density-matrix terms is smaller than that in the canonical TDDFT approach, resulting in better efficiency for the combined evaluation routine when processing large molecules.
For the remaining terms treated with the minimal auxiliary basis set, GPU memory is typically no longer a limiting factor due to the reduced size of the auxiliary space.

\subsection{FSSH Module}
We have implemented the FSSH algorithm natively in Python within the GPU4PySCF package, rather than interfacing with existing MD packages.
Our implementation follows a standard FSSH workflow \cite{Baihua2024}: initial conditions are generated via Wigner sampling, followed by an iterative loop of electronic structure calculations, nuclear propagation, and stochastic hopping.
In the following, we briefly describe the implementation.
\begin{enumerate}
    \item \textbf{Initialization:} Sample initial nuclear coordinates $\mathbf{R}_0$ and momenta $\mathbf{P}_0$ from the Wigner distribution. The quantum coefficient of the initial active state $k$ is set to $c_k=1$, with all other coefficients set to 0 ($c_j=0, \ j\neq k$).
    
    \item \textbf{Electronic Structure Calculation:} At each time step $t$, electronic structure calculations are performed to provide potential energies, nuclear forces, and derivative couplings.
    The TDDFT nuclear forces and derivative couplings are computed simultaneously within a single routine, as described in Section~\ref{sec:nacv:implementation}.
    The resulting TDDFT amplitudes and Z-vectors are retained in memory.
    They are used as initial guesses for the Casida \eqref{equ:casida} and Z-vector equations \eqref{equ:Z_vector} for the next step.
    
    To resolve the phase indeterminacy of the derivative couplings, we enforce phase consistency by tracking the sign of the wavefunction overlap between consecutive time steps.
    Various approaches have been proposed to determine the phase of derivative couplings.\cite{Pittner2009, Plasser2016, Ryabinkin2015, Lee2019, Lee2021,Chen2022}
    We estimate the overlap using an approximation based on the $\boldsymbol{X}$ component of the TDDFT amplitudes,
    \begin{align}
        \langle \Psi_{I}(t) | \Psi_{J}(t') \rangle = \sum_{ai,bj} X_{ai}^{I,t} X_{bj}^{J,t'} \langle \phi_{ai}^{t} | \phi_{bj}^{t'} \rangle,
        \label{equ:overlap}
    \end{align}
    where $\phi_{ai}^{t}$ denotes a singly excited Slater determinant corresponding to a transition from orbital $i$ to $a$.
    To compute the overlap $\langle \phi_{ai}^{t} | \phi_{bj}^{t'} \rangle$, we construct the overlap matrix between the occupied orbitals of the two Slater determinants and then take the determinant of this matrix.
    To optimize efficiency, the summation in Eq.~\eqref{equ:overlap} is truncated to include only the five dominant contributions based on the magnitude of the coefficient products $X_{ai}^{I,t} X_{bj}^{J,t'}$.
    
    \item \textbf{Nuclear Momenta Propagation (First Half-Step):} The nuclear momenta are updated by a half-step $\Delta t/2$ using the Velocity Verlet scheme:
    \begin{align}
    \mathbf{P}_{t+\Delta t/2} = \mathbf{P}_t + \mathbf{F}(\mathbf{R}_t) \cdot \frac{\Delta t}{2}.
    \end{align}
    \item \textbf{Nuclear Positions Propagation:}  the coordinates are propagated to the full time step: 
    \begin{align}
    \mathbf{R}_{t+\Delta t} = \mathbf{R}_t + \mathbf{M}^{-1} \mathbf{P}_{t+\Delta t/2} \cdot \Delta t.
    \end{align}
    
    \item\textbf{Quantum Coefficient Evolution:} The quantum coefficients are propagated using the unitary evolution operator:
    \begin{align}
    \mathbf{c}(t+\Delta t) = e^{-i\Delta t \mathbf{V}^{\text{eff}}} \mathbf{c}(t),
    \end{align}
    where the effective potential matrix $V_{kj}^{\text{eff}} = - i \hbar \dot{\mathbf{R}} \cdot \mathbf{d}_{kj}$ accounts for both the adiabatic energies and the nonadiabatic couplings.

    \item \textbf{Surface Hopping and Velocity Rescaling:} Evaluate the hopping probability $P_{k \rightarrow l}$ using the FSSH criterion as described in Section~\ref{sec:theory}. Upon a successful hop from state $k$ to $l$, the nuclear velocities must be adjusted to ensure total energy conservation. The post-hop nuclear momentum is modified along the direction of the derivative coupling:
    \begin{align}
    \mathbf{P}_{t+\Delta t/2} = \mathbf{P}_{t+\Delta t/2} - \gamma_{kl} \mathbf{d}_{kl}.
    \end{align}
    The scaling factor $\gamma_{kl}$ is determined by:
    \begin{align}
    \gamma_{kl}^2 a_{kl} - \gamma_{kl} b_{kl} - (E_{kk} - E_{ll}) = 0.
    \end{align}
    Here, $a_{kl} = \frac{1}{2} \sum_I \mathbf{d}_{I,kl}^2\cdot M_I^{-1}$ and $b_{kl} = \sum_I \mathbf{v}_I \cdot \mathbf{d}_{I,kl}$.
    When $b_{kl}^2 + 4a_{kl}(E_k - E_l) \geq 0$, the equation has real solutions, and the solution corresponding to the smaller change in velocity is selected. Otherwise, the hop is frustrated, and we set
    \begin{align}
    \gamma_{kl} = \frac{b_{kl}}{a_{kl}}.
    \end{align}

    \item \textbf{Nuclear Propagation (Second Half-Step):} After the hopping decision, the momenta are updated by the remaining half-step:
    \begin{align}
    \mathbf{P}_{t+\Delta t} = \mathbf{P}_{t+\Delta t/2} + \mathbf{F}(\mathbf{R}_{t+\Delta t}) \cdot \frac{\Delta t}{2}.
    \end{align}
    
    \item \textbf{Decoherence Correction:} To mitigate the overcoherence inherent in FSSH, decoherence correction is performed.
    Following the scheme developed by Granucci and Persico \cite{Granucci2007}, the coefficients are collapsed according to:
    \begin{align}
    c_j' = & c_j(t) e^{-\Delta t / \tau_{jk}} \quad (j \neq k),\\
    c_k' = & c_k(t) \left( \frac{1 - \sum_{j \neq k} |c_j'(t)|^2}{|c_k(t)|^2} \right)^{1/2}.
    \end{align}
    The dephasing time $\tau_{jk}$ is defined as
    \begin{equation}
        \tau_{jk} = \frac{\hbar}{|E_{jj} - E_{kk}|} \left( 1 + \frac{\alpha}{E_{\text{kin}}} \right),
    \end{equation}
    where $\alpha$ is the decoherence parameter (typically 0.1 Hartree), and $E_{\text{kin}}$ is the nuclear kinetic energy.
    
    \item \textbf{Iteration:} The process returns to Step 2 and repeats until the maximum simulation time is reached.
\end{enumerate}

\section{Performance Assessment}
\label{sec:performance}
We report the performance of derivative coupling calculations using density-fitting TDA without the minimal RI approximation, referred to as canonical TDA.
We also consider three variants of the minimal auxiliary basis approximation: the TDA-ris approximation, TDA with only the ris-Z-vector approximation (\textbf{TDA (ris-Z)}), and their combined application (\textbf{TDA-ris (ris-Z)}).
The accuracy of the approximate nuclear gradients and derivative couplings is evaluated against results obtained using canonical density-fitting TDA.

The test set consists of molecules with up to 113 atoms, including benchmark systems within the GPU4PySCF package, and two representative molecules studied in Section~\ref{sec:applications}.
To mimic realistic nonadiabatic molecular dynamics workloads, we consider a typical electronic structure step in a FSSH simulation,
in which the nuclear gradient of the first excited state is computed together with the derivative couplings between each two states, including the ground state and two lowest excited states. In these calculations, PBE0 functional\cite{adamo1999toward} and def2-TZVP\cite{weigend2005balanced} basis are used.

Performance is evaluated on NVIDIA A100-80G GPUs and RTX 4090 GPUs.
All benchmarks were performed using GPU4PySCF 1.7.0, PySCF 2.8.0, CuPy 13.4.1, and cuTENSOR 2.2.0.

\begin{table}[htp]
    \centering
    \begin{threeparttable}
    \begin{tabular}{lrrr}
    \toprule
    Molecule & TDA-ris & TDA (ris-Z) & TDA-ris (ris-Z) \\
    \midrule
Benzene       & 4.102e-04 (2.93\%)  &  1.283e-04 (0.92\%)  &  5.237e-04 (3.74\%)  \\
Vitamin C     & 7.737e-04 (3.45\%)  &  1.056e-03 (4.71\%)  &  6.786e-04 (3.03\%)  \\
Bisphenol A   & 1.907e-04 (1.47\%)  &  4.493e-04 (3.47\%)  &  5.209e-04 (4.02\%)  \\
BODIPY\tnote{a}& 2.161e-04 (0.65\%)  &  5.803e-04 (1.75\%)  &  6.882e-04 (2.07\%)  \\
Ochratoxin A  & 1.126e-03 (4.27\%)  &  4.855e-04 (1.84\%)  &  9.961e-04 (3.78\%)  \\
Cetirizine    & 4.080e-04 (1.91\%)  &  9.830e-04 (4.61\%)  &  9.496e-04 (4.46\%)  \\
Raffinose     & 6.105e-04 (3.36\%)  &  9.406e-04 (5.17\%)  &  7.949e-04 (4.37\%)  \\
TMARh\tnote{b}& 7.794e-05 (0.25\%)  &  5.658e-04 (1.81\%)  &  5.673e-04 (1.81\%)  \\
Sphingomyelin & 4.386e-05 (0.19\%)  &  1.014e-03 (4.48\%)  &  1.027e-03 (4.54\%)  \\
Azadirachtin  & 3.665e-04 (1.44\%)  &  6.657e-04 (2.62\%)  &  5.402e-04 (2.13\%)  \\
Taxol         & 4.318e-04 (2.90\%)  &  4.233e-04 (2.84\%)  &  3.114e-04 (2.09\%)  \\
\hline
Average       &  4.232e-04 (2.08\%) &  6.629e-04 (3.11\%) &  6.907e-04 (3.28\%) \\
    \bottomrule
    \end{tabular}
    \caption{RMS errors (Hartree/Bohr) in  nuclear gradients of the first excited state compared to the canonical TDA reference (relative errors in parentheses).}
    \label{tab:grad:error}
    \begin{tablenotes}[flushleft]
        \small
        \item[a] Molecular structure shown in Fig. \ref{fig:pm650structure}.
        \item[b] Molecular structure shown in Fig. \ref{fig:tmarh}.
    \end{tablenotes}
    \end{threeparttable}
\end{table}

\begin{table}[htp]
    \centering
    \begin{threeparttable}
    \begin{tabular}{lrrr}
    \toprule
    Molecule & TDA-ris & TDA (ris-Z) & TDA-ris (ris-Z) \\
    \midrule
Benzene       & 9.760e-03 (178.55\%)  &  5.606e-05 (1.03\%)  &  9.857e-03 (180.32\%)  \\
Vitamin C     & 1.936e-02 (9.36\%)  &  7.864e-03 (3.80\%)  &  2.372e-02 (11.47\%)  \\
Bisphenol A   & 1.694e-02 (5.04\%)  &  1.464e-02 (4.35\%)  &  2.445e-02 (7.27\%)  \\
BODIPY        & 1.511e-02 (4.68\%)  &  1.992e-02 (6.17\%)  &  2.553e-02 (7.92\%)  \\
Ochratoxin A  & 4.622e-01 (71.88\%)  &  2.938e-02 (4.57\%)  &  4.628e-01 (71.96\%)  \\
Cetirizine    & 4.674e-02 (5.40\%)  &  2.836e-02 (3.27\%)  &  3.518e-02 (4.06\%)  \\
Raffinose     & 1.019e-02 (33.31\%)  &  1.670e-03 (5.46\%)  &  1.138e-02 (37.19\%)  \\
TMARh         & 7.670e-04 (0.80\%)  &  7.541e-03 (7.85\%)  &  7.894e-03 (8.22\%)  \\
Sphingomyelin & 2.907e-03 (2.90\%)  &  5.107e-03 (5.09\%)  &  5.285e-03 (5.27\%)  \\
Azadirachtin  & 1.574e+00 (90.26\%)  &  1.020e-01 (5.85\%)  &  1.574e+00 (90.23\%)  \\
Taxol         & 4.399e-04 (14.17\%)  &  1.184e-04 (3.81\%)  &  4.327e-04 (13.94\%)  \\
\hline
Average       & 1.963e-01 (37.85\%) &  1.969e-02 (4.66\%) &  1.982e-01 (39.80\%) \\
    \bottomrule
    \end{tabular}
    \caption{RMS errors in derivative couplings between the first and second excited states compared to the canonical TDA reference (relative errors in parentheses).}
    \label{tab:nacv:error}
    \end{threeparttable}
\end{table}

Table~\ref{tab:grad:error} reports the RMS deviations of the first excited-state nuclear gradients for the three minimal RI approximations.
The relative error is calculated as the ratio of the RMS deviation to the norm of the gradients over all atoms.
All three variants exhibit small errors.
Their relative errors generally below 5\%.
TDA (ris-Z) and TDA-ris (ris-Z) show comparable accuracy, whereas TDA-ris is overall more accurate than the other two variants.

In Table~\ref{tab:nacv:error}, we report the errors of derivative couplings between the first and second excited states.
In this case, TDA-ris and TDA-ris (ris-Z) exhibit very similar accuracy, while TDA (ris-Z) yields smaller errors than the other two methods.
When the minimal auxiliary basis approximation is applied to the TDDFT Casida equation, both excitation energies and amplitudes are affected.
Errors in the excitation energies directly influence the derivative couplings, as these quantities appear in the denominators of the derivative coupling expressions.
Discrepancies in the excitation amplitudes impact nearly all terms in the evaluation of nonadiabatic coupling vectors.
In contrast, TDA (ris-Z) introduces errors only in the Z-vector and therefore affects fewer terms than TDA-ris.

In Table~\ref{tab:nacv:error}, large relative errors are observed for benzene, raffinose, ochratoxin A, and azadirachtin.
For benzene and raffinose, these errors can be attributed to near-zero reference values from canonical TDA calculations at the test geometries.
For ochratoxin A and azadirachtin, the first and second excited states are energetically close at the chosen geometries, with gaps on the order of 0.004 $E_\text{h}$ ($\sim$ 0.1 eV).
Because these energy differences appear in the denominators of the derivative coupling expressions, they further amplify the overall errors.

\begin{table}[htp]
    \begin{threeparttable} 
    \caption{Timings (in seconds) for SCF, TDDFT, excited state gradient and derivative coupling calculations on an NVIDIA A100 GPU.}
    \begin{tabular}{lrrrrrrrrr}
    \toprule
    & & & \multicolumn{2}{c}{Casida equation\tnote{a}} && \multicolumn{4}{c}{Gradients and derivative couplings\tnote{b}} \\
    \cline{4-5} \cline{7-10} \\
    Molecule & $N_\text{atoms}$ & SCF & TD & TDA-ris && \shortstack{TDA\\canonical} & \shortstack{TDA\\(ris-Z)} & \shortstack{TDA-ris\\~~} & \shortstack{TDA-ris\\(ris-Z)} \\
    \midrule
    Benzene          & 12   & 0.54   & 0.55   & 0.11   && 1.68   & 1.41   & 1.78   & 1.38   \\
    Vitamin C        & 20   & 1.12   & 1.26   & 0.19   && 4.92   & 3.97   & 4.50   & 3.46   \\
    Bisphenol A      & 33   & 1.79   & 2.53   & 0.38   && 10.61  & 7.81   & 9.20   & 6.96   \\
    BODIPY\tnote{c}  & 40   & 3.13   & 4.26   & 0.62   && 19.25  & 14.34  & 17.33  & 12.20  \\
    Ochratoxin A     & 45   & 5.53   & 9.93   & 1.47   && 25.65  & 18.60  & 23.99  & 16.00  \\
    Cetirizine       & 52   & 5.60   & 10.02  & 1.29   && 29.19  & 21.07  & 27.42  & 17.86  \\
    Raffinose        & 66   & 9.14   & 30.58  & 2.84   && 50.40  & 33.86  & 46.65  & 28.53  \\
    TMARh\tnote{d}   & 73   & 13.8   & 22.82  & 2.61   && 71.42  & 44.54  & 65.63  & 37.42  \\
    Sphingomyelin    & 84   & 11.72  & 15.95  & 3.11   && 47.97  & 31.57  & 45.23  & 26.42  \\
    Azadirachtin     & 95   & 85.83  & 127.89 & 9.53   && 240.79 & 112.27 & 223.43 & 90.64  \\
    Taxol            & 113  & 145.34 & 210.48 & 13.33  && 430.78 & 199.41 & 410.95 & 154.20 \\
    \bottomrule
    \end{tabular}
    \begin{tablenotes}[flushleft] 
        \small  
        \item[a] In total 5 lowest excited states are calculated.
        \item[b] Including the nuclear gradients of the first excited state, and 6 derivative couplings corresponding to all pairs among the 3 lowest excited states.
        \item[c] Molecular structure is shown in Fig. \ref{fig:pm650structure}.
        \item[d] Molecular structure is shown in Fig. \ref{fig:tmarh}.
    \end{tablenotes}
    \label{tab:gpu_performance_A100}
    \end{threeparttable}
\end{table}

\begin{table}[htp]
    \begin{threeparttable}
    \caption{Timings (in seconds) for SCF, TDDFT, excited state gradient and derivative coupling calculations on an NVIDIA RTX 4090 GPU}
    \begin{tabular}{lrrrrrrrrr}
    \toprule
    & & & \multicolumn{2}{c}{Casida equation\tnote{a}} && \multicolumn{4}{c}{Gradients and derivative couplings\tnote{b}} \\
    \cline{4-5} \cline{7-10} \\
    Molecule & $N_\text{atoms}$ & SCF & TD & TDA-ris && \shortstack{TDA\\canonical} & \shortstack{TDA\\(ris-Z)} & \shortstack{TDA-ris\\~~} & \shortstack{TDA-ris\\(ris-Z)} \\
    \midrule
    Benzene         & 12   & 1.68   & 1.22   & 1.05   && 4.76   & 4.73   & 5.18   & 4.24    \\
    Vitamin C       & 20   & 4.33   & 6.86   & 1.18   && 18.24  & 13.16  & 16.71  & 10.72   \\
    Bisphenol A     & 33   & 8.13   & 18.7   & 1.46   && 49.35  & 32.29  & 44.13  & 25.48   \\
    BODIPY\tnote{c} & 40   & 18.16  & 38.32  & 1.71   && 104.47 & 60.17  & 94.48  & 47.12   \\
    Ochratoxin A    & 45   & 32.28  & 96.48  & 2.67   && 171.37 & 89.85  & 154.32 & 68.52   \\
    Cetirizine      & 52   & 37.47  & 103.4  & 2.84   && 205.93 & 105.58 & 184.19 & 79.59   \\
    Raffinose       & 66   & 84.54  & 329.58 & 5.31   && 422.8  & 212.54 & 367.07 & 154.35  \\
    TMARh\tnote{d}  & 73   & 125.29 & 252.5  & 5.35   && 621.61 & 290.96 & 562.3  & 209.41  \\
    \bottomrule
    \end{tabular}
    \begin{tablenotes}[flushleft] 
        \small 
        \item[a] In total 5 lowest excited states are calculated.
        \item[b] Including the nuclear gradients of the first excited state, and 6 derivative couplings corresponding to all pairs among the 3 lowest excited states.
        \item[c] Molecular structure is shown in Fig. \ref{fig:pm650structure}.
        \item[d] Molecular structure is shown in Fig. \ref{fig:tmarh}.
    \end{tablenotes}
    
    \label{tab:gpu_performance_4090}
    \end{threeparttable}
\end{table}

Table~\ref{tab:gpu_performance_A100} summarizes the computational cost for SCF calculations and excited-state calculations on an NVIDIA A100 GPU.
The overall performance (including solving the Casida equation, computing nuclear gradients, and evaluating derivative couplings) of TDA (ris-Z) and TDA-ris is comparable.
In TDA-ris, the cost of solving the Casida equation is negligible, whereas in TDA (ris-Z) the cost of solving the Z-vector equations is negligible. The remaining computational steps have similar costs in the two approaches, leading to comparable overall performance.
Both methods are approximately 30\% - 50\% faster than canonical TDA.
Approximating both the Casida and Z-vector equations provides an additional $\sim$50\% speed-up, making TDA-ris (ris-Z) about 2 - 3 times faster than canonical TDA.
It should be noted that the performance differences between canonical TDA and TDA-ris observed here are smaller than those reported in our previous study\cite{Pu2026}.
This change is primarily due to the merging of multiple computationally demanding two-electron integral evaluation steps.
As the cost of the two-electron integrals decreases, the benefit of using a minimal auxiliary basis set for the Z-vector and integral derivatives becomes less pronounced.

The advantages of the minimal auxiliary basis approximation become more pronounced for large molecules, particularly for systems approaching 100 atoms.
In such systems, the evaluation of derivative couplings must be divided into small batches due to GPU memory limitations.
Repeated evaluation of the expensive two-electron integrals introduces additional overhead.
By using a minimal auxiliary basis set, the GPU memory requirement is reduced.
It reduces the number of batches as well as the overhead of repeated two-electron integral evaluations.

Compared to ground-state SCF, canonical TDA excited-state calculations require roughly 5 - 10 times more computational time.
Approximately, 30\% of the excited-state computation time is spent on the Casida equation.
Applying the ris approximation to the Casida equation makes its computational cost negligible.
When combined with the Z-vector approximation, TDA-ris (ris-Z) further lowers the excited-state computational cost, bringing it close to that of a ground-state calculation.

The SCF and excited-state calculation timings on the NVIDIA RTX 4090 GPU are summarized in Table~\ref{tab:gpu_performance_4090}.
The overall performance on the RTX 4090 GPU is approximately 5 - 10 times slower than on the A100 GPU.
The advantages of using a minimal auxiliary basis set are more significant on the RTX 4090 GPU.
For larger systems, canonical TDDFT calculations on the RTX 4090 are approximately 10 times slower than on the A100 GPU.
This performance gap is reduced to about 6-fold when ris approximations are employed.
Since density fitting is used to evaluate two-electron integrals, the main computational cost in these calculations arises from the associated tensor contraction operations.
Compared to the A100 GPU, the 4090 GPU has significantly lower double-precision performance for tensor contractions.
The minimal auxiliary basis approximation becomes particularly effective on the 4090 GPU as it reduces the cost of tensor contraction.
As a result, the TDA-ris (ris-Z) approach is roughly 4 times faster than canonical TDA.
TDA (ris-Z) achieves approximately a 50\% speed-up to canonical TDA.
TDA-ris is slightly faster than TDA (ris-Z) by about 20\% in most systems.

\section{Applications}
\label{sec:applications}
We present three example systems to demonstrate the capabilities of our FSSH implementation:
\begin{itemize}
\item \textbf{Benzene molecule:} The dynamics on the low-lying states of the benzene molecule, serving as a
test system to validate the correctness of our implementation.
\item \textbf{BODIPY derivative:} The excited-state dynamics of the first singlet state of a
boron-dipyrromethene (BODIPY) molecule\cite{Joo2021}, representing a medium-sized system that
can be efficiently modeled using the GPU4PySCF TDDFT features.
\item \textbf{TMARh molecular system:} A more complex tetramethylamino rhodamine (TMARh) molecular system\cite{Zhang2022}, which
is typically considered computationally expensive and slow when simulated using CPU-based programs.
\end{itemize}

\subsection{Benchmark Validation: Internal Conversion in Benzene}
To rigorously validate our FSSH implementation, we investigated the ultrafast internal conversion (IC) of benzene in vacuum. The low-lying excited states of benzene offer a diverse range of electronic characteristics: the $S_1$ and $S_2$ states arise from $\pi \to \pi^*$ transitions, where $S_1$ is symmetry-forbidden (dominated by $\text{HOMO–1}\to\text{LUMO}$ excitation) and $S_2$ is symmetry-allowed (dominated by  $\text{HOMO}\to\text{LUMO}$); $S_3$ is an allowed $\sigma \to \pi^*$ transition (dominated by  $\text{HOMO–2}\to\text{LUMO}$). These well-defined interstate characteristics make benzene an ideal benchmark for testing nonadiabatic dynamics simulation methods. Our analysis focuses on the population transfer among the $S_1$, $S_2$, and $S_3$ manifolds.

For the initial conditions, we sampled 100 geometries from a Wigner distribution at 300 K to account for nuclear thermal fluctuations.
Electronic structure calculations included four lowest excited states ($S_1$-$S_4$) to avoid artificial truncation of potential relaxation channels. We employed the PBE0 functional with the def2-SVP basis set under the TDA approximation to analytically compute energies, gradients, and derivative couplings. A total of 100 independent trajectories were propagated for 250 fs with the time step being 0.5 fs, all initiated from the $S_3$ state ($c_3 = 1$). To address the over-coherence issue, the energy-based decoherence correction proposed by Granucci and Persico\cite{Granucci2007} was employed with the parameter $\alpha = 0.1$ Hartree. 

\begin{figure}[htbp]
    \centering
    \includegraphics[width=0.75\linewidth]{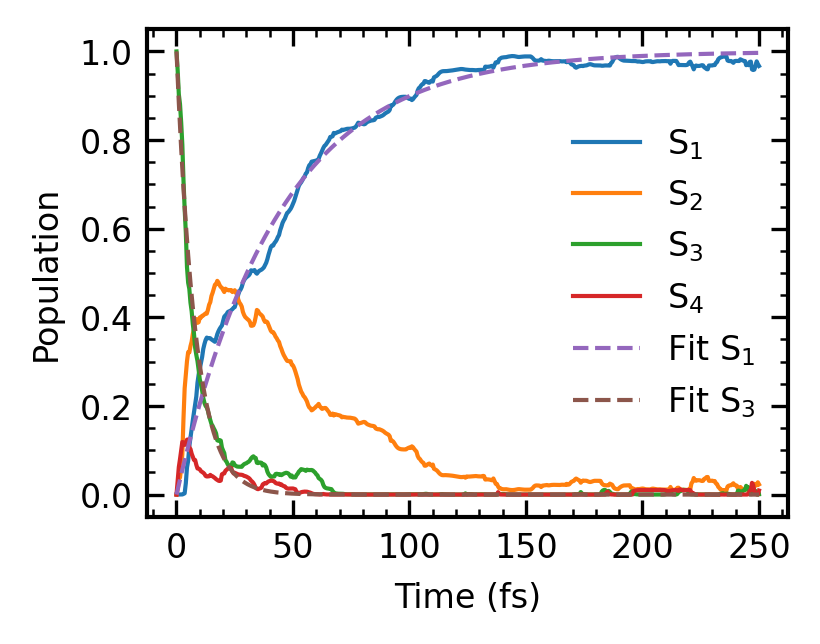}
    \caption{The population of each state of benzene against simulation time}
    \label{fig:benzene_pop}
\end{figure}

As illustrated in Figure \ref{fig:benzene_pop}, benzene exhibits a characteristic cascade relaxation following excitation to $S_3$. The $S_3$ population decays rapidly to near zero within 50 fs, indicating an ultrafast electronic quenching.  The $S_2$ population peaks at approximately 25 fs before further relaxing toward $S_1$ via regions of strong nonadiabatic coupling. Eventually, the population stabilizes on the $S_1$ surface. To quantify these dynamics, the decay of $S_3$ and the rise of $S_1$ were fitted to single-exponential models:
\begin{equation}
    p_3\approx e^{-t/\tau_3},
\end{equation}
\begin{equation}
    p_1\approx 1-e^{-t/\tau_1}.
\end{equation}
Our fitted $S_3$ lifetime ($\tau_3 = 8.2$ fs) is in excellent agreement with the NWChem simulations by Song et al. (11.2 fs)\cite{Song2020}, confirming the reliability of our implementation. However, a discrepancy is observed in the $S_1$ rise time (43.6 fs in this work vs. 57.1 fs in the reference\cite{Song2020}). We attribute this to the non-adiabatic coupling evaluation method. While the reference utilized wavefunction overlap approximations, we employed a rigorous analytical gradient approach. Such approximations are often sufficient for ultrafast transitions like $S_3 \to S_2$, numerical errors tend to accumulate during slower relaxation processes like $S_2 \to S_1$, leading to deviations over longer timescales. This comparison underscores the importance of using analytically computed derivative couplings for capturing long-term internal conversion.
\begin{figure}[htbp]
    \centering
    \includegraphics[width=0.75\linewidth]{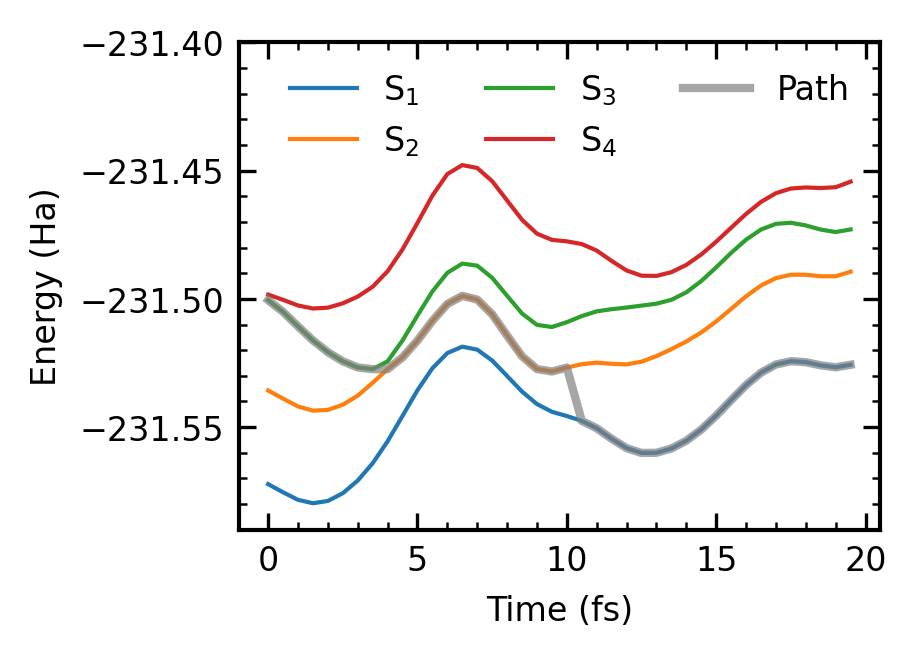}
    \caption{FSSH trajectory of benzene showing relaxation from the third excited state (S3) to the first excited state (S1)}
    \label{fig:benzene_traj}
\end{figure}

Figure \ref{fig:benzene_traj} depicts a representative trajectory. Around 4 fs, the potential energy surfaces of $S_3$ and $S_2$ become nearly degenerate, accompanied by a spike in NAC magnitude, which triggers an instantaneous transition. This behavior is consistent with the short $S_3$ lifetime derived from the ensemble-averaged population. Notably, an $S_2 \to S_1$ transition occurs around 10.5 fs, where the energy gap between the two states remains substantial. Despite non-negligible adiabatic energy gap at this geometry between the two states, non-negligible non-adiabatic coupling exists in this geometric region due to nuclear motion, causing the hopping probability to accumulate gradually and eventually exceed the threshold. This example highlights that non-adiabatic transitions can be driven by strong coupling even when the adiabatic energy gap is large, a feature captured correctly by the analytical derivative coupling used in this work.




\subsection{Excited-State Dynamics of BODIPY: Efficiency and Validation}
\begin{figure}[htbp]
    \centering
    \includegraphics[width=0.75\linewidth]{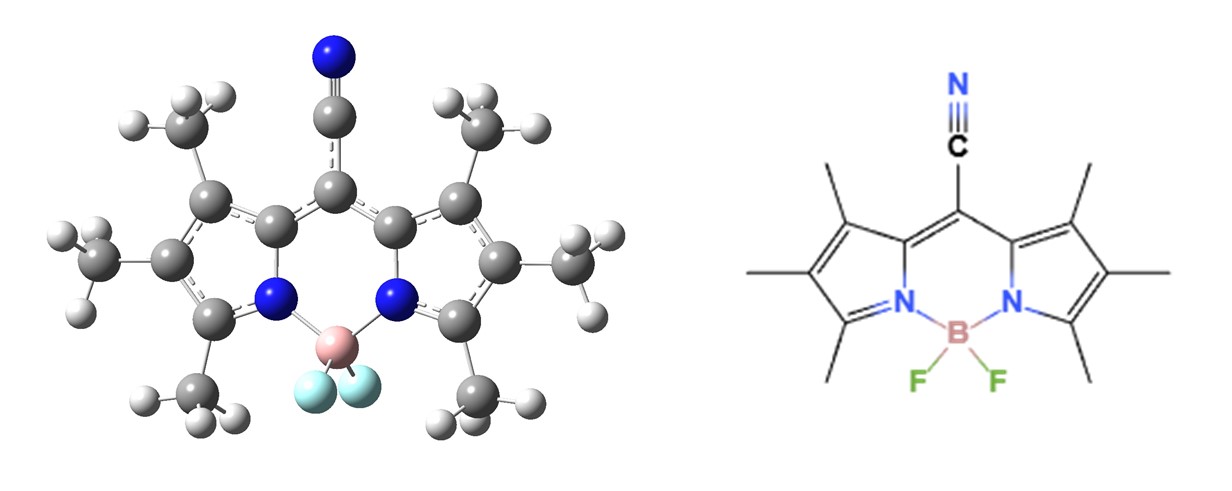}
    \caption{Structure of PM650}
    \label{fig:pm650structure}
\end{figure}

To evaluate the practical performance of our implementation on chemically complex systems, we investigated a boron-dipyrromethene (BODIPY) derivative, PM650 (structure shown in Fig. \ref{fig:pm650structure}). BODIPY-based fluorophores are distinguished by their exceptional photophysical properties, including high molar extinction coefficients\cite{Luo2014}, near-unity fluorescence quantum yields\cite{Watanabe2026,José2024}, and narrow emission bands\cite{Ma2022}. These characteristics have led to their widespread application in bioimaging\cite{Kumar2025}, light-harvesting systems\cite{Yu2025}, and photovoltaic devices\cite{nie2025}. Experimentally, femtosecond pump-probe spectroscopy has shown that upon excitation to the second singlet state ($S_2$), PM650 undergoes ultrafast internal conversion to the $S_1$ state in less than 20 fs \cite{Joo2021}.

50 initial conditions were sampled from a Wigner distribution at 300 K based on the ground-state equilibrium geometry. Given the ultrafast nature of the $S_2 \to S_1$ transition, we employed a total simulation time of 60 fs with a small time step of 0.2 fs. A total of 50 independent trajectories were propagated, with electronic structures calculated at the CAM-B3LYP/def2-SVP level of theory. To ensure physical consistency in the slower relaxation components, the energy-based decoherence correction with $\alpha = 0.1$ Hartree was applied.

\begin{figure}[htbp]
    \centering
    \includegraphics[width=0.75\linewidth]{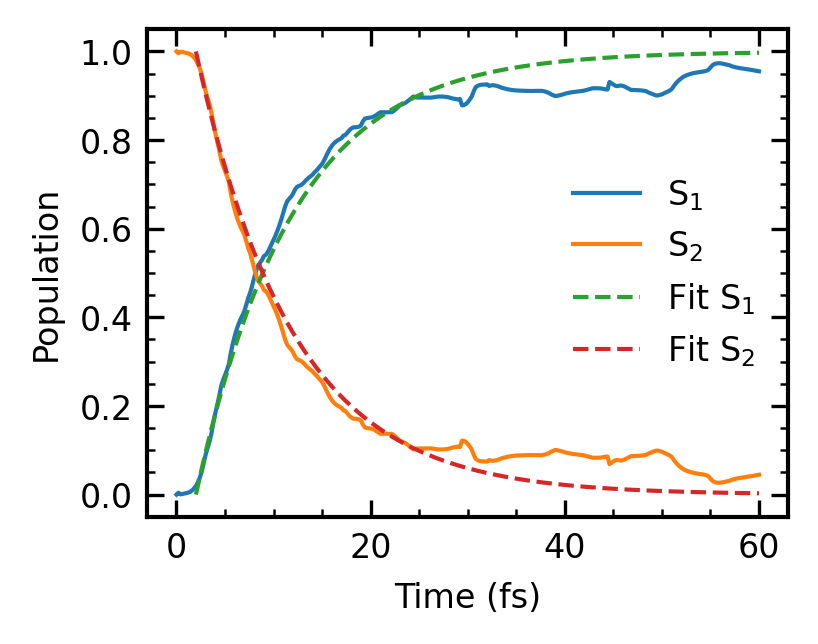}
    \caption{The population of S1 and S2 of PM650 against simulation time using canonical TDA}
    \label{fig:bodipy_fssh}
\end{figure}

As shown in Figure \ref{fig:bodipy_fssh}, the $S_2$ population decays precipitously, reaching near zero within 50 fs. A single-exponential fit of the population evolution yields an IC time constant of 10.9 fs, which is in excellent agreement with the experimental upper bound ($< 20$ fs). Furthermore, our result align closely with high-level wavepacket dynamics studies by Neethu et al., who reported a time constant of 9.6 fs using the multiconfiguration time-dependent Hartree (MCTDH) method \cite{Joo2025}. This high degree of concordance validates the reliability of our TDDFT-based FSSH approach in capturing complex nonadiabatic dynamics, offering a robust balance between accuracy and computational efficiency compared to wavepacket-based methods.

\begin{figure}[htbp]
    \centering
    \includegraphics[width=0.75\linewidth]{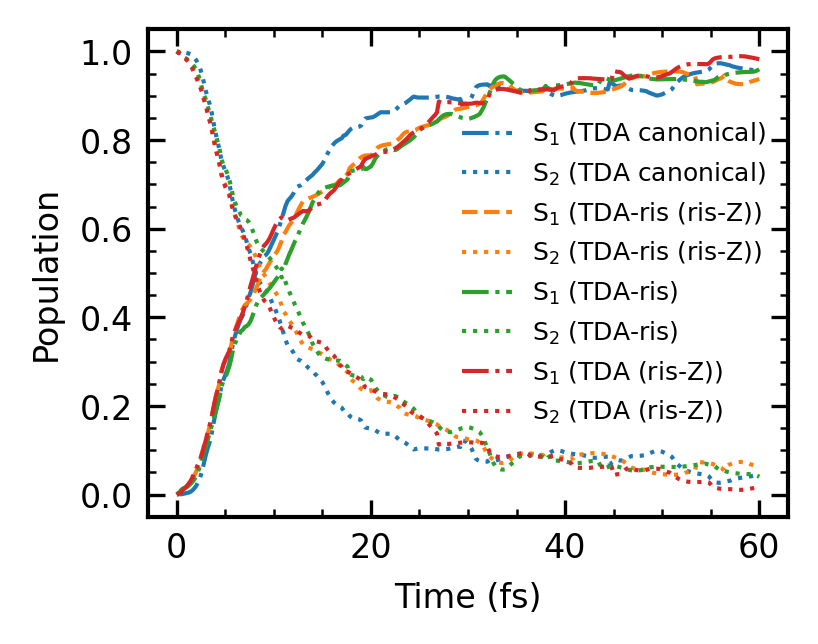}
    \caption{The population of the first excited state (S1) and the second excited state (S2) of PM650 against simulation time using exact and approximate TDDFT methods.}
    \label{fig:bodipy_ris}
\end{figure}

Beyond the canonical implementation, we evaluated the performance of several acceleration schemes, including TDA-ris, TDA (ris-Z), and TDA-ris (ris-Z), on an NVIDIA A100 GPU platform. The improvements in computational performance resulted from the minimal auxiliary basis set approximation are consistent with the benchmarks reported in Section~\ref{sec:performance}.
Specifically, the TDA-ris (ris-Z) approach achieves a 1.7-fold speed-up over the canonical TDA reference, while TDA-ris and TDA (ris-Z) yield 1.4-fold and 1.3-fold speed-ups, respectively.
Importantly, these approximations do not compromise the physical integrity of the simulated dynamics.
As shown in Figure \ref{fig:bodipy_ris}, the state populations computed with the approximate methods give IC time constants of 12.9 fs (TDA-ris (ris-Z)), 13.8 fs (TDA-ris), and 12.7 fs (TDA (ris-Z)), deviating by a maximum of 2.9 fs from the canonical TDA value of 10.9 fs. This demonstrates that the ris-based approximations provide a high-fidelity alternative for large-scale NAMD simulations, significantly reducing the wall-clock time without altering the predicted kinetic mechanisms.

\subsection{Large-Scale Application: Nonadiabatic Dynamics of TMARh}

Building upon the successful validation of the benzene and BODIPY models, we extended our framework to the tetramethylamino rhodamine (TMARh) molecular system. 
TMARh serves as a prototypical “off-on” pH-sensitive probe.\cite{Best2010,Sørensen2016} Thus, elucidating the microscopic mechanisms governing its fluorescence quenching is vital for the rational design of next-generation chemical sensors. Experimental studies using femtosecond transient absorption spectroscopy have revealed that upon photoexcitation to $S_2$ state, TMARh undergoes IC to $S_1$ state within approximately 100 fs \cite{Zhang2022}.
\begin{figure}[htbp]
    \centering
    \includegraphics[width=0.75\linewidth]{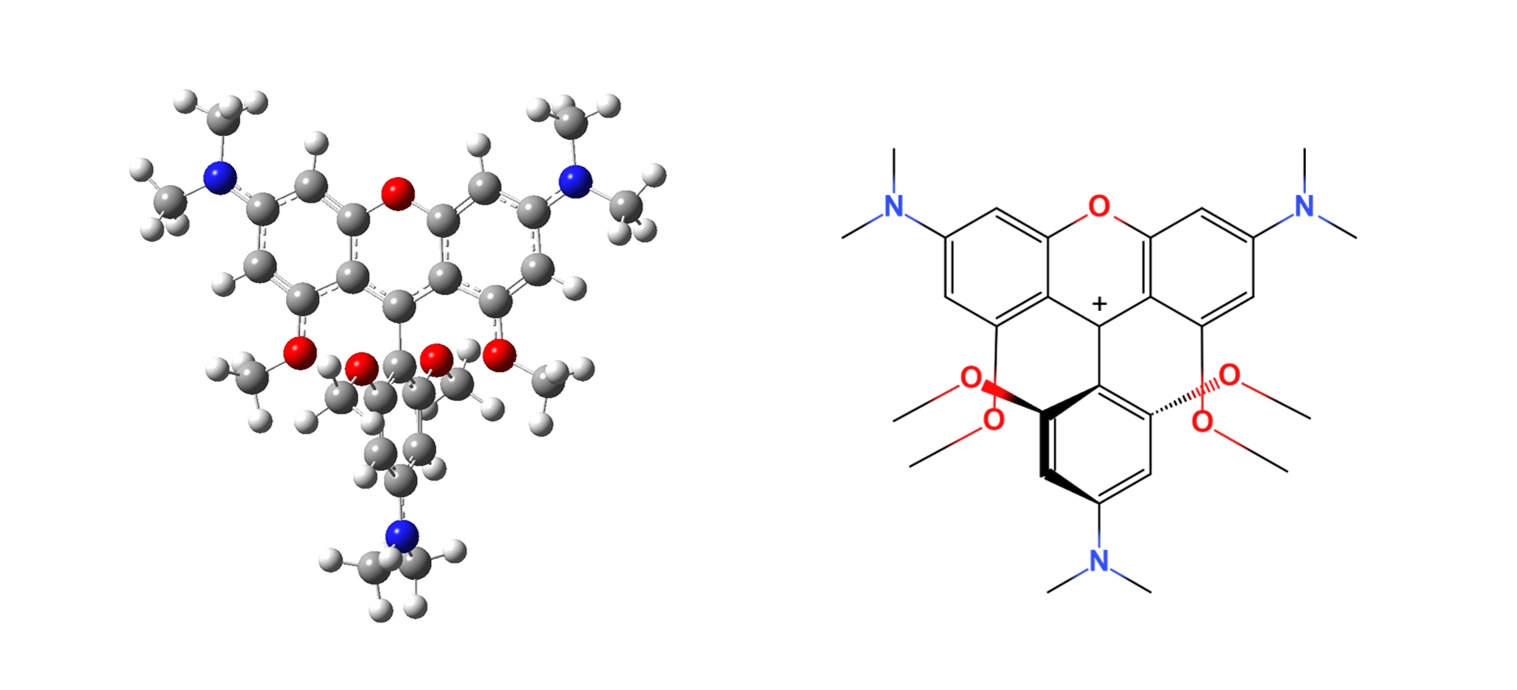}
    \caption{Structure of TMARh}
    \label{fig:tmarh}
\end{figure}

\begin{figure}[htbp]
    \centering
    \includegraphics[width=0.75\linewidth]{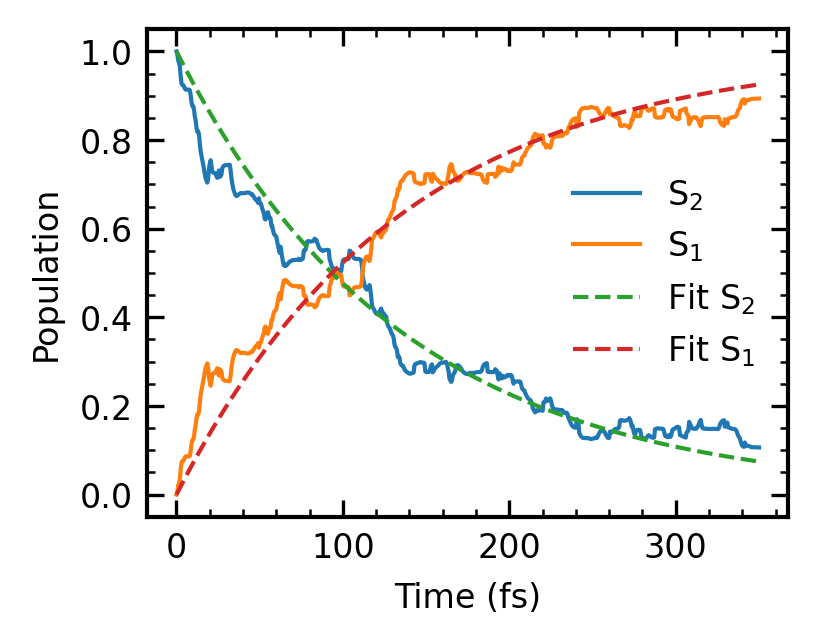}
    \caption{The population of the first excited state ($S_1$) and the second excited state ($S_2$) of TMARh against simulation time using the def2-TZVP basis set}
    \label{fig:tmarh_tzvp}
\end{figure}

\begin{figure}
    \centering
    \includegraphics[width=0.75\linewidth]{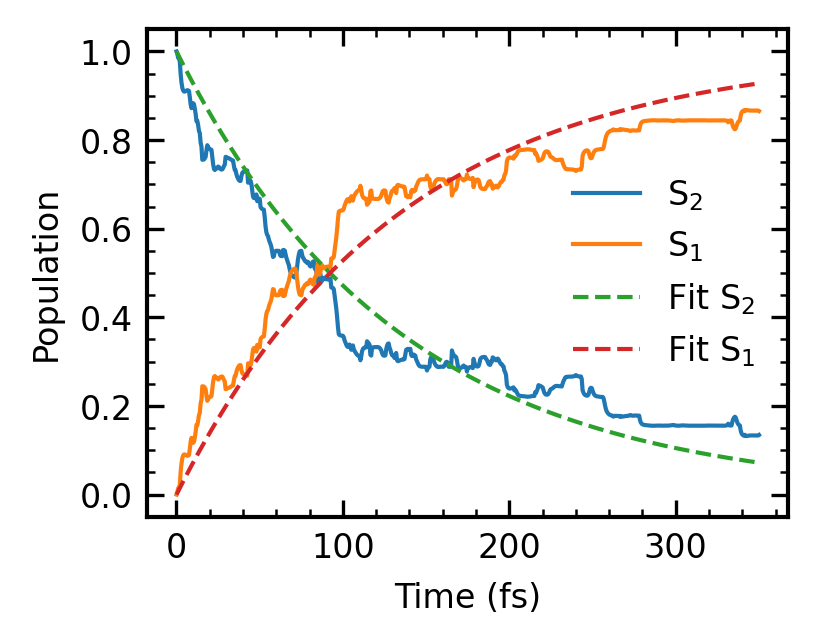}
    \caption{The population of the first excited state ($S_1$) and the second excited state ($S_2$) of TMARh against simulation time using the def2-SVP basis set}
    \label{fig:tmarh_svp}
\end{figure}
To ensure the quantitative reliability of our dynamics, we first investigated the sensitivity of the relaxation kinetics to the atomic basis set. Simulations were conducted using both the def2-SVP and def2-TZVP basis sets, with 50 trajectories initiated from the $S_2$ state and propagated for 350 fs using 0.5 fs time step. The energy-based decoherence correction was applied with $\alpha = 0.1$ Hartree .
As shown in Fig. \ref{fig:tmarh_tzvp} and \ref{fig:tmarh_svp}, the IC time constant computed with the triple-zeta def2-TZVP basis set is 134.8 fs, which aligns well with the experimental $\sim$100 fs regime. Interestingly, the double-zeta def2-SVP basis set yields a nearly identical time constant of 133.1 fs. This close agreement suggests that the excited-state potential energy surfaces and the corresponding nonadiabatic couplings are well-converged at the def2-SVP level for this system.

\begin{figure}[htbp]
    \centering
    \includegraphics[width=0.75\linewidth]{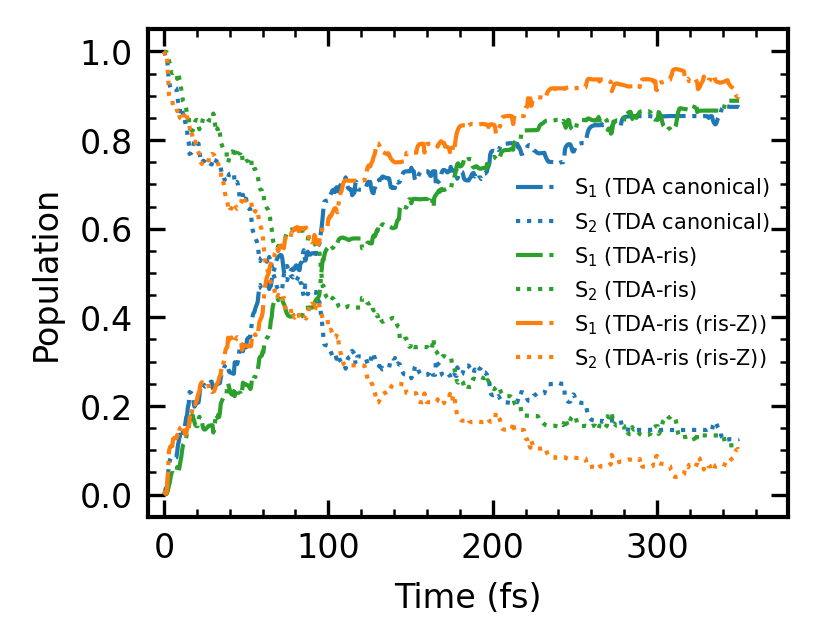}
    \caption{The population of the first excited state ($S_1$) and the second excited state ($S_2$) of TMARh against simulation time using TDDFT-ris methods}
    \label{fig:tmarh:ris}
\end{figure}

We further assessed the performance of our ris-based acceleration schemes on this large-scale system. The population decay curves for canonical TDA, TDA-ris, and TDA-ris (ris-Z) are compared in Fig. \ref{fig:tmarh:ris}, employing the same simulation protocols as the def2-SVP basis set tests.
The trajectories obtained with the approximate methods exhibit remarkable overlap with the canonical TDA reference, all yielding a consistent IC time constant of approximately 105.0 fs.

The computational speed-up observed for TMARh follows the same encouraging trend as seen in the BODIPY case. On the NVIDIA A100 platform, the TDA-ris (ris-Z) scheme provides the highest efficiency, achieving a 1.6-fold speed-up over the canonical TDA implementation, while the TDA-ris offers a 1.3-fold improvement.

\section{Conclusion}
In this work, we extend our previous development of TDDFT-ris nuclear gradients and derivative couplings to fewest-switches surface hopping (FSSH) nonadiabatic dynamics simulations.
The GPU4PySCF package has been extended to support FSSH dynamics, providing a flexible framework for integrating excited-state electronic structure methods within PySCF and GPU4PySCF.

The minimal auxiliary basis set approximation (ris) can be applied to both the Casida equation and the Z-vector equations in the evaluation of excited-state nuclear gradients and derivative couplings.
The ris approximation for the Z-vector equations (ris-Z) yields small errors to canonical TDDFT.
Its relative errors in both gradients and derivative couplings are generally within 10\%, which is sufficient for most FSSH simulations.
The TDDFT-ris method and the combined TDDFT-ris (ris-Z) approach exhibit somewhat larger errors.
However, these do not lead to significant deviations in practical FSSH dynamics.
Benchmark FSSH simulations confirm that these ris approximations introduce negligible errors, with the resulting trajectories essentially identical to those obtained using canonical TDDFT.

In terms of computational performance, the Z-vector approximation alone provides an approximately 30\% speed-up over canonical TDDFT.
The TDDFT-ris and ris-Z approaches exhibit comparable efficiency, while the combined TDDFT-ris (ris-Z) approach offers the best overall performance.
The TDDFT-ris (ris-Z) approach provides speedups of up to 2.8$\times$ on an NVIDIA A100 GPU (relative to canonical TDDFT calculation) and up to 3.0$\times$ on an RTX 4090 GPU.
In our current implementation, A100 GPUs are better suited for practical nonadiabatic dynamics simulations.
Compared to RTX 4090 GPUs, it offers approximately 6$\times$ speedups for ris approximations and 10$\times$ for canonical TDDFT calculations.
By employing the GPU-accelerated TDDFT with minimal auxiliary basis approximations as the electronic structure engine, FSSH simulations for systems with up to 73 atoms using a triple-$\zeta$ basis set can achieve more than 1500 simulation steps per day on a single NVIDIA A100 GPU.

\section{Acknowledgement}
The authors thank Baihua Wu, Yunlong Xiao, and Tai Wang for valuable discussions on the implementation of the FSSH algorithm and the selection of test cases. The authors would also like to thank the ByteDance Volcano Engine cloud and Shenzhen Bay Laboratory Supercomputing Centre for providing computational resources.

\bibliographystyle{plainnat}
\bibliography{references} 

@article{Tully1990,
    author = {Tully, John C.},
    title = {Molecular dynamics with electronic transitions},
    journal = {The Journal of Chemical Physics},
    volume = {93},
    number = {2},
    pages = {1061-1071},
    year = {1990},
    month = {07},
    issn = {0021-9606},
    doi = {10.1063/1.459170},
    url = {https://doi.org/10.1063/1.459170},
    eprint = {https://pubs.aip.org/aip/jcp/article-pdf/93/2/1061/18987588/1061_1_online.pdf},
}

@article{pyxaid,
author = {Akimov, Alexey V. and Prezhdo, Oleg V.},
title = {The PYXAID Program for Non-Adiabatic Molecular Dynamics in Condensed Matter Systems},
journal = {Journal of Chemical Theory and Computation},
volume = {9},
number = {11},
pages = {4959-4972},
year = {2013},
doi = {10.1021/ct400641n},
    note ={PMID: 26583414},

URL = { 
        https://doi.org/10.1021/ct400641n
},
eprint = { 
        https://doi.org/10.1021/ct400641n
}
}

@article{SHARC,
author = {Mai, Sebastian and Marquetand, Philipp and González, Leticia},
title = {Nonadiabatic dynamics: The SHARC approach},
journal = {WIREs Computational Molecular Science},
volume = {8},
number = {6},
pages = {e1370},
doi = {https://doi.org/10.1002/wcms.1370},
url = {https://wires.onlinelibrary.wiley.com/doi/abs/10.1002/wcms.1370},
eprint = {https://wires.onlinelibrary.wiley.com/doi/pdf/10.1002/wcms.1370},
year = {2018}
}

@article{hefeinamd,
    author = {Zheng, Qijing and Chu, Weibin and Zhao, Chuanyu and Zhang, Lili and Guo, Hongli and Wang, Yanan and Jiang, Xiang and Zhao, Jin},
    author+an = {1=first; 8=corresponding,highlight},
    doi = {10.1002/wcms.1411},
    issn = {17590884},
    journal = {Wiley Interdisciplinary Reviews: Computational Molecular Science},
    month = {March},
    number = {6},
    pages = {e1411},
    publisher = {Wiley Periodicals, Inc.},
    title = {Ab Initio Nonadiabatic Molecular Dynamics Investigations on the Excited Carriers in Condensed Matter Systems},
    volume = {9},
    year = {2019}
}

@article{newtonx,
author = {Barbatti, Mario and Bondanza, Mattia and Crespo-Otero, Rachel and Demoulin, Baptiste and Dral, Pavlo O. and Granucci, Giovanni and Kossoski, F{\'a}bris and Lischka, Hans and Mennucci, Benedetta and Mukherjee, Saikat and Pederzoli, Marek and Persico, Maurizio and Pinheiro Jr, Max and Pittner, Ji{\v{r}}{\'i} and Plasser, Felix and Sangiogo Gil, Eduarda and Stojanovic, Ljiljana},
title = {Newton-X Platform: New Software Developments for Surface Hopping and Nuclear Ensembles},
journal = {Journal of Chemical Theory and Computation},
volume = {18},
number = {11},
pages = {6851-6865},
year = {2022},
doi = {10.1021/acs.jctc.2c00804},
    note ={PMID: 36194696},
URL = { 
        https://doi.org/10.1021/acs.jctc.2c00804
},
eprint = { 
        https://doi.org/10.1021/acs.jctc.2c00804
}
}

@Article{pyrai2md,
author ="Li, Jingbai and Reiser, Patrick and Boswell, Benjamin R. and Eberhard, André and Burns, Noah Z. and Friederich, Pascal and Lopez, Steven A.",
title  ="Automatic discovery of photoisomerization mechanisms with nanosecond machine learning photodynamics simulations",
journal  ="Chem. Sci.",
year  ="2021",
volume  ="12",
issue  ="14",
pages  ="5302-5314",
publisher  ="The Royal Society of Chemistry",
doi  ="10.1039/D0SC05610C",
url  ="http://dx.doi.org/10.1039/D0SC05610C",
}

@article{jadenamd,
author = {Du, Likai and Lan, Zhenggang},
title = {An On-the-Fly Surface-Hopping Program JADE for Nonadiabatic Molecular Dynamics of Polyatomic Systems: Implementation and Applications},
journal = {Journal of Chemical Theory and Computation},
volume = {11},
number = {4},
pages = {1360-1374},
year = {2015},
doi = {10.1021/ct501106d},
    note ={PMID: 26574348},
URL = {    
        https://doi.org/10.1021/ct501106d
},
eprint = { 
        https://doi.org/10.1021/ct501106d
}
}

@Article{furche2011,
author ="Tapavicza, Enrico and Meyer, Alexander M. and Furche, Filipp",
title  ="Unravelling the details of vitamin D photosynthesis by non-adiabatic molecular dynamics simulations",
journal  ="Phys. Chem. Chem. Phys.",
year  ="2011",
volume  ="13",
issue  ="47",
pages  ="20986-20998",
publisher  ="The Royal Society of Chemistry",
doi  ="10.1039/C1CP21292C",
url  ="http://dx.doi.org/10.1039/C1CP21292C",
}

@article{Myers2024,
    author = {Myers, Christopher A. and Miyazaki, Ken and Trepl, Thomas and Isborn, Christine M. and Ananth, Nandini},
    title = {GPU-accelerated on-the-fly nonadiabatic semiclassical dynamics},
    journal = {The Journal of Chemical Physics},
    volume = {161},
    number = {8},
    pages = {084114},
    year = {2024},
    month = {08},
    issn = {0021-9606},
    doi = {10.1063/5.0223628},
    url = {https://doi.org/10.1063/5.0223628},
    eprint = {https://pubs.aip.org/aip/jcp/article-pdf/doi/10.1063/5.0223628/20131977/084114_1_5.0223628.pdf},
}

@Article{Tully1998,
author ="C. Tully, John",
title  ="Mixed quantum–classical dynamics",
journal  ="Faraday Discuss.",
year  ="1998",
volume  ="110",
issue  ="0",
pages  ="407-419",
publisher  ="The Royal Society of Chemistry",
doi  ="10.1039/A801824C",
url  ="http://dx.doi.org/10.1039/A801824C",
}

@article{Peters2019,
author = {Peters, Laurens D. M. and Kussmann, J{\"o}rg and Ochsenfeld, Christian},
title = {Nonadiabatic Molecular Dynamics on Graphics Processing Units: Performance and Application to Rotary Molecular Motors},
journal = {Journal of Chemical Theory and Computation},
volume = {15},
number = {12},
pages = {6647-6659},
year = {2019},
doi = {10.1021/acs.jctc.9b00859},
    note ={PMID: 31763834},
URL = {   
        https://doi.org/10.1021/acs.jctc.9b00859
},
eprint = { 
        https://doi.org/10.1021/acs.jctc.9b00859
}
}

@article{Peters2020,
author = {Peters, Laurens D. M. and Kussmann, J{\"o}rg and Ochsenfeld, Christian},
title = {Combining Graphics Processing Units, Simplified Time-Dependent Density Functional Theory, and Finite-Difference Couplings to Accelerate Nonadiabatic Molecular Dynamics},
journal = {The Journal of Physical Chemistry Letters},
volume = {11},
number = {10},
pages = {3955-3961},
year = {2020},
doi = {10.1021/acs.jpclett.0c00320},
    note ={PMID: 32374606},
URL = { 
        https://doi.org/10.1021/acs.jpclett.0c00320
},
eprint = { 
        https://doi.org/10.1021/acs.jpclett.0c00320
}

}

@misc{Wu2024,
      title={Enhancing GPU-acceleration in the Python-based Simulations of Chemistry Framework},
      author={Xiaojie Wu and Qiming Sun and Zhichen Pu and Tianze Zheng and Wenzhi Ma and Wen Yan and Xia Yu and Zhengxiao Wu and Mian Huo and Xiang Li and Weiluo Ren and Sheng Gong and Yumin Zhang and Weihao Gao},
      year={2024},
      eprint={2404.09452},
      archivePrefix={arXiv},
      primaryClass={physics.comp-ph},
      url={https://arxiv.org/abs/2404.09452},
}

@article{Pu2026,
author = {Pu, Zhichen and Wu, Xiaojie and Wang, Yuanheng and Fan, Cheng and Yan, Wen and Zhou, Zehao and Gao, Yi Qin and Sun, Qiming},
title = {Analytical Excited-State Gradients and Derivative Couplings in TDDFT with Minimal Auxiliary Basis Set Approximation and GPU Acceleration},
journal = {Journal of Chemical Theory and Computation},
volume = {0},
number = {0},
pages = {null},
year = {0},
doi = {10.1021/acs.jctc.5c01960},
    note ={PMID: 41615830},
URL = { 
        https://doi.org/10.1021/acs.jctc.5c01960
},
eprint = { 
        https://doi.org/10.1021/acs.jctc.5c01960
}
}

@article{Zhou2023,
author = {Zhou, Zehao and Della Sala, Fabio and Parker, Shane M.},
title = {Minimal Auxiliary Basis Set Approach for the Electronic Excitation Spectra of Organic Molecules},
journal = {The Journal of Physical Chemistry Letters},
volume = {14},
number = {7},
pages = {1968-1976},
year = {2023},
doi = {10.1021/acs.jpclett.2c03698},
    note ={PMID: 36787711},
URL = { 
        https://doi.org/10.1021/acs.jpclett.2c03698
},
eprint = { 
        https://doi.org/10.1021/acs.jpclett.2c03698
}
}

@article{Martin2005,
author = {Dreuw, Andreas and Head-Gordon, Martin},
title = {Single-Reference ab Initio Methods for the Calculation of Excited States of Large Molecules},
journal = {Chemical Reviews},
volume = {105},
number = {11},
pages = {4009-4037},
year = {2005},
doi = {10.1021/cr0505627},
    note ={PMID: 16277369},
URL = { 
        https://doi.org/10.1021/cr0505627
},
eprint = { 
        https://doi.org/10.1021/cr0505627
}

}

@article{Furche2002,
    author = {Furche, Filipp and Ahlrichs, Reinhart},
    title = {Adiabatic time-dependent density functional methods for excited state properties},
    journal = {The Journal of Chemical Physics},
    volume = {117},
    number = {16},
    pages = {7433-7447},
    year = {2002},
    month = {10},
    issn = {0021-9606},
    doi = {10.1063/1.1508368},
    url = {https://doi.org/10.1063/1.1508368},
    eprint = {https://pubs.aip.org/aip/jcp/article-pdf/117/16/7433/19317804/7433_1_online.pdf},
}

@article{Handy1984,
    author = {Handy, Nicholas C. and Schaefer, Henry F., III},
    title = {On the evaluation of analytic energy derivatives for correlated wave functions},
    journal = {The Journal of Chemical Physics},
    volume = {81},
    number = {11},
    pages = {5031-5033},
    year = {1984},
    month = {12},
    issn = {0021-9606},
    doi = {10.1063/1.447489},
    url = {https://doi.org/10.1063/1.447489},
    eprint = {https://pubs.aip.org/aip/jcp/article-pdf/81/11/5031/18950442/5031_1_online.pdf},
}

@misc{wu2025designingQCalgorithms,
      title={Designing quantum chemistry algorithms with just-in-time compilation}, 
      author={Xiaojie Wu, Qiming Sun and Yuanheng Wang},
      year={2025},
      eprint={2507.09772},
      archivePrefix={arXiv},
      primaryClass={physics.comp-ph},
      url={https://arxiv.org/abs/2507.09772}, 
}

@article{Baihua2024,
author = {Wu, Baihua and He, Xin and Liu, Jian},
title = {Nonadiabatic Field on Quantum Phase Space: A Century after Ehrenfest},
journal = {The Journal of Physical Chemistry Letters},
volume = {15},
number = {2},
pages = {644-658},
year = {2024},
doi = {10.1021/acs.jpclett.3c03385},
    note ={PMID: 38205956},
URL = { 
        https://doi.org/10.1021/acs.jpclett.3c03385
},
eprint = { 
        https://doi.org/10.1021/acs.jpclett.3c03385
}
}

@article{Granucci2007,
    author = {Granucci, Giovanni and Persico, Maurizio},
    title = {Critical appraisal of the fewest switches algorithm for surface hopping},
    journal = {The Journal of Chemical Physics},
    volume = {126},
    number = {13},
    pages = {134114},
    year = {2007},
    month = {04},
    issn = {0021-9606},
    doi = {10.1063/1.2715585},
    url = {https://doi.org/10.1063/1.2715585},
    eprint = {https://pubs.aip.org/aip/jcp/article-pdf/doi/10.1063/1.2715585/15397070/134114_1_online.pdf},
}

@article{Pittner2009,
title = {Optimization of mixed quantum-classical dynamics: Time-derivative coupling terms and selected couplings},
journal = {Chemical Physics},
volume = {356},
number = {1},
pages = {147-152},
year = {2009},
note = {Moving Frontiers in Quantum Chemistry:},
issn = {0301-0104},
doi = {https://doi.org/10.1016/j.chemphys.2008.10.013},
url = {https://www.sciencedirect.com/science/article/pii/S0301010408004783},
author = {Jiri Pittner and Hans Lischka and Mario Barbatti},
}

@article{Plasser2016,
author = {Plasser, Felix and Ruckenbauer, Matthias and Mai, Sebastian and Oppel, Markus and Marquetand, Philipp and González, Leticia},
title = {Efficient and Flexible Computation of Many-Electron Wave Function Overlaps},
journal = {Journal of Chemical Theory and Computation},
volume = {12},
number = {3},
pages = {1207-1219},
year = {2016},
doi = {10.1021/acs.jctc.5b01148},
    note ={PMID: 26854874},
URL = { 
        https://doi.org/10.1021/acs.jctc.5b01148
},
eprint = { 
        https://doi.org/10.1021/acs.jctc.5b01148
}
}

@article{Ryabinkin2015,
author = {Ryabinkin, Ilya G. and Nagesh, Jayashree and Izmaylov, Artur F.},
title = {Fast Numerical Evaluation of Time-Derivative Nonadiabatic Couplings for Mixed Quantum-Classical Methods},
journal = {The Journal of Physical Chemistry Letters},
volume = {6},
number = {21},
pages = {4200-4203},
year = {2015},
doi = {10.1021/acs.jpclett.5b02062},
    note ={PMID: 26538034},
URL = { 
        https://doi.org/10.1021/acs.jpclett.5b02062
},
eprint = { 
        https://doi.org/10.1021/acs.jpclett.5b02062
}
}

@article{Lee2019,
author = {Lee, Seunghoon and Kim, Eunji and Lee, Sangyoub and Choi, Cheol Ho},
title = {Fast Overlap Evaluations for Nonadiabatic Molecular Dynamics Simulations: Applications to SF-TDDFT and TDDFT},
journal = {Journal of Chemical Theory and Computation},
volume = {15},
number = {2},
pages = {882-891},
year = {2019},
doi = {10.1021/acs.jctc.8b01049},
URL = { 
        https://doi.org/10.1021/acs.jctc.8b01049
},
eprint = { 
        https://doi.org/10.1021/acs.jctc.8b01049
}
}

@article{Lee2021,
author = {Lee, Seunghoon and Horbatenko, Yevhen and Filatov, Michael and Choi, Cheol Ho},
title = {Fast and Accurate Computation of Nonadiabatic Coupling Matrix Elements Using the Truncated Leibniz Formula and Mixed-Reference Spin-Flip Time-Dependent Density Functional Theory},
journal = {The Journal of Physical Chemistry Letters},
volume = {12},
number = {19},
pages = {4722-4728},
year = {2021},
doi = {10.1021/acs.jpclett.1c00932},
    note ={PMID: 33983029},
URL = { 
        https://doi.org/10.1021/acs.jpclett.1c00932
},
eprint = { 
        https://doi.org/10.1021/acs.jpclett.1c00932
}
}

@article{Song2020,
author = {Song, Huajing and Fischer, Sean A. and Zhang, Yu and Cramer, Christopher J. and Mukamel, Shaul and Govind, Niranjan and Tretiak, Sergei},
title = {First Principles Nonadiabatic Excited-State Molecular Dynamics in NWChem},
journal = {Journal of Chemical Theory and Computation},
volume = {16},
number = {10},
pages = {6418-6427},
year = {2020},
doi = {10.1021/acs.jctc.0c00295},
    note ={PMID: 32808780},
URL = { 
        https://doi.org/10.1021/acs.jctc.0c00295
},
eprint = { 
        https://doi.org/10.1021/acs.jctc.0c00295
}
}

@Article{Joo2021,
author ="Lee, Changmin and Seo, Kiho and Kim, Munnyon and Joo, Taiha",
title  ="Coherent internal conversion from high lying electronic states to S1 in boron-dipyrromethene derivatives",
journal  ="Phys. Chem. Chem. Phys.",
year  ="2021",
volume  ="23",
issue  ="44",
pages  ="25200-25209",
publisher  ="The Royal Society of Chemistry",
doi  ="10.1039/D1CP03513D",
url  ="http://dx.doi.org/10.1039/D1CP03513D",
}

@Article{Joo2025,
author ="Anand, Neethu and Kim, Munnyon and Lee, Changmin and Ma, Jinhyuk and Joo, Taiha",
title  ="Understanding contrasting S2 → S1 internal conversion rates in boron-dipyrromethene derivatives via multi-configuration time-dependent hartree method",
journal  ="Phys. Chem. Chem. Phys.",
year  ="2025",
volume  ="27",
issue  ="44",
pages  ="23550-23560",
publisher  ="The Royal Society of Chemistry",
doi  ="10.1039/D5CP02771C",
url  ="http://dx.doi.org/10.1039/D5CP02771C",
}

@Article{Watanabe2026,
  author   = {Watanabe, Keita and Honda, Gentaro and Terauchi, Yuki and Mamiya, Shunsuke and Inaba, Yuya and Nakajima, Tasuku and Gong, Jian Ping and Yamaguchi, Yusaku and Kitagawa, Yuichi and Hasegawa, Yasuchika and Ide, Yuki and Gao, Min and Yoneda, Tomoki and Inokuma, Yasuhide},
  title    = {Superacid-resistant macrocyclic BODIPYs},
  journal  = {Nature Communications},
  year     = {2026},
  volume   = {17},
  number   = {1},
  pages    = {2332},
  month    = mar,
  doi      = {10.1038/s41467-026-70499-9},
  url      = {https://doi.org/10.1038/s41467-026-70499-9},
  issn     = {2041-1723},
}

@article{Luo2014,
  author  = {Luo, Liang and Wu, Di and Li, Wei and Zhang, Shuai and Ma, Yuanhong and Yan, Su and You, Jingsong},
  title   = {Regioselective Decarboxylative Direct C–H Arylation of Boron Dipyrromethenes (BODIPYs) at 2,6-Positions: A Facile Access to a Diversity-Oriented BODIPY Library},
  journal = {Organic Letters},
  year    = {2014},
  volume  = {16},
  number  = {23},
  pages   = {6080-6083},
  doi     = {10.1021/ol502883x}
}

@article{José2024,
title = {From blue to red. Reaching the full visible spectrum with a single fluorophore: BODIPY},
journal = {Tetrahedron},
volume = {168},
pages = {134334},
year = {2024},
issn = {0040-4020},
doi = {https://doi.org/10.1016/j.tet.2024.134334},
url = {https://www.sciencedirect.com/science/article/pii/S0040402024005155},
author = {José G. Becerra-González and Eduardo Peña-Cabrera and José L. Belmonte-Vázquez},
}

@article{Ma2022,
title = {Creation of BODIPYs-based red OLEDs with high color purity via modulating the energy gap and restricting rotation of substituents},
journal = {Dyes and Pigments},
volume = {203},
pages = {110377},
year = {2022},
issn = {0143-7208},
doi = {https://doi.org/10.1016/j.dyepig.2022.110377},
url = {https://www.sciencedirect.com/science/article/pii/S0143720822002996},
author = {Daiyu Ma and Guimin Zhao and Haowen Chen and Renyin Zhou and Guanghao Zhang and Wenwen Tian and Wei Jiang and Yueming Sun},
}

@Article{Kumar2025,
AUTHOR = {Kumar, Panangattukara Prabhakaran Praveen and Saxena, Shivanjali and Joshi, Rakesh},
TITLE = {BODIPY Dyes: A New Frontier in Cellular Imaging and Theragnostic Applications},
JOURNAL = {Colorants},
VOLUME = {4},
YEAR = {2025},
NUMBER = {2},
ARTICLE-NUMBER = {13},
URL = {https://www.mdpi.com/2079-6447/4/2/13},
ISSN = {2079-6447},
DOI = {10.3390/colorants4020013}
}

@article{Yu2025,
  author  = {Yu, Longyue and Huang, Xionghui and Feng, Ning and Fu, Wenwen and Xin, Xia and Hao, Jingcheng and Li, Hongguang},
  title   = {Solvent-Free Artificial Light-Harvesting System in a Fluid Donor with Highly Efficient F\"{o}rster Resonance Energy Transfer},
  journal = {The Journal of Physical Chemistry Letters},
  year    = {2025},
  volume  = {16},
  number  = {5},
  pages   = {1305-1311},
  doi     = {10.1021/acs.jpclett.4c03518}
}

@article{nie2025,
title = {Improving photovoltaic performance of organic solar cells utilizing medium bandgap BODIPY non-fused acceptor as third component via ternary strategy},
journal = {Journal of Molecular Structure},
volume = {1335},
pages = {141977},
year = {2025},
issn = {0022-2860},
doi = {https://doi.org/10.1016/j.molstruc.2025.141977},
url = {https://www.sciencedirect.com/science/article/pii/S0022286025006623},
author = {Ziqi Nie and Tingting Gu and Xu Liang and Rahul Singhal and Haijun Xu and Ganesh D. Sharma},
}

@article{Sørensen2016,
author = {Sørensen, Thomas Just and Shi, Dong and Laursen, Bo W.},
title = {Tetramethoxy-Aminorhodamine (TMARh): A Bichromophore, an Improved Fluorophore, and a pH Switch},
journal = {Chemistry – A European Journal},
volume = {22},
number = {21},
pages = {7046-7049},
doi = {https://doi.org/10.1002/chem.201600496},
url = {https://chemistry-europe.onlinelibrary.wiley.com/doi/abs/10.1002/chem.201600496},
eprint = {https://chemistry-europe.onlinelibrary.wiley.com/doi/pdf/10.1002/chem.201600496},
year = {2016}
}

@article{Best2010,
  author  = {Best, Quinn A. and Xu, Ruisong and McCarroll, Matthew E. and Wang, Lichang and Dyer, Daniel J.},
  title   = {Design and Investigation of a Series of Rhodamine-Based Fluorescent Probes for Optical Measurements of pH},
  journal = {Organic Letters},
  year    = {2010},
  volume  = {12},
  number  = {14},
  pages   = {3219-3221},
  doi     = {10.1021/ol1011967}
}

@Article{Zhang2022,
author ="Zhang, Wei and Zhao, Li and Laursen, Bo W. and Chen, Junsheng",
title  ="Revealing the sensing mechanism of a fluorescent pH probe based on a bichromophore approach",
journal  ="Phys. Chem. Chem. Phys.",
year  ="2022",
volume  ="24",
issue  ="43",
pages  ="26731-26737",
publisher  ="The Royal Society of Chemistry",
doi  ="10.1039/D2CP04339D",
url  ="http://dx.doi.org/10.1039/D2CP04339D",
}

@article{Nelson2020,
author = {Nelson, Tammie R. and White, Alexander J. and Bjorgaard, Josiah A. and Sifain, Andrew E. and Zhang, Yu and Nebgen, Benjamin and Fernandez-Alberti, Sebastian and Mozyrsky, Dmitry and Roitberg, Adrian E. and Tretiak, Sergei},
title = {Non-adiabatic Excited-State Molecular Dynamics: Theory and Applications for Modeling Photophysics in Extended Molecular Materials},
journal = {Chemical Reviews},
volume = {120},
number = {4},
pages = {2215-2287},
year = {2020},
doi = {10.1021/acs.chemrev.9b00447},
note ={PMID: 32040312},
URL = {  https://doi.org/10.1021/acs.chemrev.9b00447 },
eprint = {  https://doi.org/10.1021/acs.chemrev.9b00447 }
}

@Article{Tapavicza2013,
author ="Tapavicza, Enrico and Bellchambers, Gregory D. and Vincent, Jordan C. and Furche, Filipp",
title  ="Ab initio non-adiabatic molecular dynamics",
journal  ="Phys. Chem. Chem. Phys.",
year  ="2013",
volume  ="15",
issue  ="42",
pages  ="18336-18348",
publisher  ="The Royal Society of Chemistry",
doi  ="10.1039/C3CP51514A",
url  ="http://dx.doi.org/10.1039/C3CP51514A"}

@article{Parker2017,
author = {Parker, Shane M. and Rappoport, Dmitrij and Furche, Filipp},
title = {Quadratic Response Properties from TDDFT: Trials and Tribulations},
journal = {Journal of Chemical Theory and Computation},
volume = {14},
number = {2},
pages = {807-819},
year = {2018},
doi = {10.1021/acs.jctc.7b01008},
    note ={PMID: 29232511},
URL = {  https://doi.org/10.1021/acs.jctc.7b01008 },
eprint = {  https://doi.org/10.1021/acs.jctc.7b01008 }
}

@article{Wang2020,
author = {Wang, Linjun and Qiu, Jing and Bai, Xin and Xu, Jiabo},
title = {Surface hopping methods for nonadiabatic dynamics in extended systems},
journal = {WIREs Computational Molecular Science},
volume = {10},
number = {2},
pages = {e1435},
doi = {https://doi.org/10.1002/wcms.1435},
url = {https://wires.onlinelibrary.wiley.com/doi/abs/10.1002/wcms.1435},
eprint = {https://wires.onlinelibrary.wiley.com/doi/pdf/10.1002/wcms.1435},
year = {2020}
}

@article{Crespo-Otero2018,
author = {Crespo-Otero, Rachel and Barbatti, Mario},
title = {Recent Advances and Perspectives on Nonadiabatic Mixed Quantum–Classical Dynamics},
journal = {Chemical Reviews},
volume = {118},
number = {15},
pages = {7026-7068},
year = {2018},
doi = {10.1021/acs.chemrev.7b00577},
    note ={PMID: 29767966},
URL = {  https://doi.org/10.1021/acs.chemrev.7b00577 },
eprint = {  https://doi.org/10.1021/acs.chemrev.7b00577 }
}

@article{Qiu2023,
author = {Qiu, Tian and Climent, Clàudia and Subotnik, Joseph E.},
title = {A Practical Approach to Wave Function Propagation, Hopping Probabilities, and Time Steps in Surface Hopping Calculations},
journal = {Journal of Chemical Theory and Computation},
volume = {19},
number = {10},
pages = {2744-2757},
year = {2023},
doi = {10.1021/acs.jctc.3c00126},
    note ={PMID: 37130302},
URL = {  https://doi.org/10.1021/acs.jctc.3c00126 },
eprint = {  https://doi.org/10.1021/acs.jctc.3c00126 }
}

@article{Peng2022,
author = {Peng, Wei-Tao and Brey, Dominik and Giannini, Samuele and Dell’Angelo, David and Burghardt, Irene and Blumberger, Jochen},
title = {Exciton Dissociation in a Model Organic Interface: Excitonic State-Based Surface Hopping versus Multiconfigurational Time-Dependent Hartree},
journal = {The Journal of Physical Chemistry Letters},
volume = {13},
number = {31},
pages = {7105-7112},
year = {2022},
doi = {10.1021/acs.jpclett.2c01928},
    note ={PMID: 35900333},
URL = {  https://doi.org/10.1021/acs.jpclett.2c01928 },
eprint = {  https://doi.org/10.1021/acs.jpclett.2c01928 }
}

@article{Vogt2025,
author = {Vogt, Jan-Robert and Schulz, Michael and Souza Mattos, Rafael and Barbatti, Mario and Persico, Maurizio and Granucci, Giovanni and Hutter, J{\"u}rg and Hehn, Anna},
title = {A Density Functional Theory and Semiempirical Framework for Trajectory Surface Hopping on Extended Systems},
journal = {Journal of Chemical Theory and Computation},
volume = {21},
number = {20},
pages = {10474-10488},
year = {2025},
doi = {10.1021/acs.jctc.5c01082},
    note ={PMID: 41108029},
URL = {  https://doi.org/10.1021/acs.jctc.5c01082 },
eprint = {  https://doi.org/10.1021/acs.jctc.5c01082 }
}

@article{Chen2022,
author = {Chen, Hsing-Ta and Chen, Junhan and Cofer-Shabica, D. Vale and Zhou, Zeyu and Athavale, Vishikh and Medders, Gregory and Menger, Maximilian F. S. J. and Subotnik, Joseph E. and Jin, Zuxin},
title = {Methods to Calculate Electronic Excited-State Dynamics for Molecules on Large Metal Clusters with Many States: Ensuring Fast Overlap Calculations and a Robust Choice of Phase},
journal = {Journal of Chemical Theory and Computation},
volume = {18},
number = {6},
pages = {3296-3307},
year = {2022},
doi = {10.1021/acs.jctc.1c01304},
    note ={PMID: 35609255},
URL = {  https://doi.org/10.1021/acs.jctc.1c01304 },
eprint = {  https://doi.org/10.1021/acs.jctc.1c01304 }
}

@incollection{casida1995time,
  title={Time-dependent density functional response theory for molecules},
  author={Casida, Mark E},
  booktitle={Recent Advances In Density Functional Methods: (Part I)},
  pages={155--192},
  year={1995},
  publisher={World Scientific}
}

@article{Casida2009,
title = {Time-dependent density-functional theory for molecules and molecular solids},
journal = {Journal of Molecular Structure: THEOCHEM},
volume = {914},
number = {1},
pages = {3-18},
year = {2009},
note = {Time-dependent density-functional theory for molecules and molecular solids},
issn = {0166-1280},
doi = {https://doi.org/10.1016/j.theochem.2009.08.018},
url = {https://www.sciencedirect.com/science/article/pii/S0166128009005363},
author = {Mark E. Casida},
}

@article{Barbatti2011,
author = {Barbatti, Mario},
title = {Nonadiabatic dynamics with trajectory surface hopping method},
journal = {WIREs Computational Molecular Science},
volume = {1},
number = {4},
pages = {620-633},
doi = {https://doi.org/10.1002/wcms.64},
url = {https://wires.onlinelibrary.wiley.com/doi/abs/10.1002/wcms.64},
eprint = {https://wires.onlinelibrary.wiley.com/doi/pdf/10.1002/wcms.64},
year = {2011}
}

@article{tully2012perspective,
  title={Perspective: Nonadiabatic dynamics theory},
  author={Tully, John C},
  journal={The Journal of chemical physics},
  volume={137},
  number={22},
  year={2012},
  publisher={AIP Publishing}
}

@article{Sun2020,
    author = {Sun, Qiming and Zhang, Xing and Banerjee, Samragni and Bao, Peng and Barbry, Marc and Blunt, Nick S. and Bogdanov, Nikolay A. and Booth, George H. and Chen, Jia and Cui, Zhi-Hao and Eriksen, Janus J. and Gao, Yang and Guo, Sheng and Hermann, Jan and Hermes, Matthew R. and Koh, Kevin and Koval, Peter and Lehtola, Susi and Li, Zhendong and Liu, Junzi and Mardirossian, Narbe and McClain, James D. and Motta, Mario and Mussard, Bastien and Pham, Hung Q. and Pulkin, Artem and Purwanto, Wirawan and Robinson, Paul J. and Ronca, Enrico and Sayfutyarova, Elvira R. and Scheurer, Maximilian and Schurkus, Henry F. and Smith, James E. T. and Sun, Chong and Sun, Shi-Ning and Upadhyay, Shiv and Wagner, Lucas K. and Wang, Xiao and White, Alec and Whitfield, James Daniel and Williamson, Mark J. and Wouters, Sebastian and Yang, Jun and Yu, Jason M. and Zhu, Tianyu and Berkelbach, Timothy C. and Sharma, Sandeep and Sokolov, Alexander Yu. and Chan, Garnet Kin-Lic},
    title = {Recent developments in the PySCF program package},
    journal = {The Journal of Chemical Physics},
    volume = {153},
    number = {2},
    pages = {024109},
    year = {2020},
    month = {07},
    issn = {0021-9606},
    doi = {10.1063/5.0006074},
    url = {https://doi.org/10.1063/5.0006074},
    eprint = {https://pubs.aip.org/aip/jcp/article-pdf/doi/10.1063/5.0006074/16722275/024109_1_online.pdf},
}

@article{rappoport2005analytical,
  title={Analytical time-dependent density functional derivative methods within the RI-J approximation, an approach to excited states of large molecules},
  author={Rappoport, Dmitrij and Furche, Filipp},
  journal={The Journal of chemical physics},
  volume={122},
  number={6},
  year={2005},
  publisher={AIP Publishing}
}

@article{scalmani2006geometries,
  title={Geometries and properties of excited states in the gas phase and in solution: Theory and application of a time-dependent density functional theory polarizable continuum model},
  author={Scalmani, Giovanni and Frisch, Michael J and Mennucci, Benedetta and Tomasi, Jacopo and Cammi, Roberto and Barone, Vincenzo},
  journal={The Journal of chemical physics},
  volume={124},
  number={9},
  year={2006},
  publisher={AIP Publishing}
}

@article{tapavicza2007trajectory,
  title={Trajectory Surface Hopping within Linear Response Time-Dependent Density-Functional Theory},
  author={Tapavicza, Enrico and Tavernelli, Ivano and Rothlisberger, Ursula},
  journal={Physical review letters},
  volume={98},
  number={2},
  pages={023001},
  year={2007},
  publisher={APS}
}

@article{herbert2016beyond,
  title={Beyond time-dependent density functional theory using only single excitations: Methods for computational studies of excited states in complex systems},
  author={Herbert, John M and Zhang, Xing and Morrison, Adrian F and Liu, Jie},
  journal={Accounts of chemical research},
  volume={49},
  number={5},
  pages={931--941},
  year={2016},
  publisher={ACS Publications}
}

@article{zhang2014analytic,
  title={Analytic derivative couplings for spin-flip configuration interaction singles and spin-flip time-dependent density functional theory},
  author={Zhang, Xing and Herbert, John M},
  journal={The Journal of Chemical Physics},
  volume={141},
  number={6},
  year={2014},
  publisher={AIP Publishing}
}

@article{zhang2015analytic, title={Analytic derivative couplings in time-dependent density functional theory: Quadratic response theory versus pseudo-wavefunction approach},
  author={Zhang, Xing and Herbert, John M},
  journal={The Journal of chemical physics},
  volume={142},
  number={6},
  year={2015},
  publisher={AIP Publishing}
}

@article{send2010first,
  title={First-order nonadiabatic couplings from time-dependent hybrid density functional response theory: Consistent formalism, implementation, and performance},
  author={Send, Robert and Furche, Filipp},
  journal={The Journal of chemical physics},
  volume={132},
  number={4},
  year={2010},
  publisher={AIP Publishing}
}

@article{li2014first,
  title={First order nonadiabatic coupling matrix elements between excited states: Implementation and application at the TD-DFT and pp-TDA levels},
  author={Li, Zhendong and Suo, Bingbing and Liu, Wenjian},
  journal={The Journal of chemical physics},
  volume={141},
  number={24},
  year={2014},
  publisher={AIP Publishing}
}

@article{li2014first2,
  title={First-order nonadiabatic coupling matrix elements between excited states: A Lagrangian formulation at the CIS, RPA, TD-HF, and TD-DFT levels},
  author={Li, Zhendong and Liu, Wenjian},
  journal={The Journal of chemical physics},
  volume={141},
  number={1},
  year={2014},
  publisher={AIP Publishing}
}

@article{subotnik2015requisite,
  title={The requisite electronic structure theory to describe photoexcited nonadiabatic dynamics: Nonadiabatic derivative couplings and diabatic electronic couplings},
  author={Subotnik, Joseph E and Alguire, Ethan C and Ou, Qi and Landry, Brian R and Fatehi, Shervin},
  journal={Accounts of chemical research},
  volume={48},
  number={5},
  pages={1340--1350},
  year={2015},
  publisher={ACS Publications}
}

@article{ou2015first,
  title={First-order derivative couplings between excited states from adiabatic TDDFT response theory},
  author={Ou, Qi and Bellchambers, Gregory D and Furche, Filipp and Subotnik, Joseph E},
  journal={The Journal of chemical physics},
  volume={142},
  number={6},
  year={2015},
  publisher={AIP Publishing}
}

@article{zhang2021nonadiabatic,
  title={Nonadiabatic dynamics with spin-flip vs linear-response time-dependent density functional theory: A case study for the protonated Schiff base C5H6NH2+},
  author={Zhang, Xing and Herbert, John M},
  journal={The Journal of Chemical Physics},
  volume={155},
  number={12},
  year={2021},
  publisher={AIP Publishing}
}

@article{zhou2024converging,
  title={Converging Time-Dependent Density Functional Theory Calculations in Five Iterations with Minimal Auxiliary Preconditioning},
  author={Zhou, Zehao and Parker, Shane M},
  journal={Journal of Chemical Theory and Computation},
  volume={20},
  number={15},
  pages={6738--6746},
  year={2024},
  publisher={ACS Publications}
}

@article{li2025introducing,
  title={Introducing GPU acceleration into the python-based simulations of chemistry framework},
  author={Li, Rui and Sun, Qiming and Zhang, Xing and Chan, Garnet Kin-Lic},
  journal={The Journal of Physical Chemistry A},
  volume={129},
  number={5},
  pages={1459--1468},
  year={2025},
  publisher={ACS Publications}
}

@article{adamo1999toward,
  title={Toward reliable density functional methods without adjustable parameters: The PBE0 model},
  author={Adamo, Carlo and Barone, Vincenzo},
  journal={The Journal of chemical physics},
  volume={110},
  number={13},
  pages={6158--6170},
  year={1999},
  publisher={American Institute of Physics}
}

@article{weigend2005balanced,
  title={Balanced basis sets of split valence, triple zeta valence and quadruple zeta valence quality for H to Rn: Design and assessment of accuracy},
  author={Weigend, Florian and Ahlrichs, Reinhart},
  journal={Physical Chemistry Chemical Physics},
  volume={7},
  number={18},
  pages={3297--3305},
  year={2005},
  publisher={Royal Society of Chemistry}
}

@misc{nvidia_nvcc,
  author       = {{NVIDIA Corporation}},
  title        = {CUDA Compiler Driver NVCC Documentation – Just-In-Time Compilation},
  year         = {2026},
  howpublished = {\url{https://docs.nvidia.com/cuda/cuda-compiler-driver-nvcc/#just-in-time-compilation}},
  note         = {Accessed: 2026-04-02}
}

@misc{nvidia_cutensor,
  author       = {{NVIDIA Corporation}},
  title        = {cuTENSOR Documentation – Just-In-Time Compilation},
  year         = {2026},
  howpublished = {\url{https://docs.nvidia.com/cuda/cutensor/latest/just_in_time_compilation.html}},
  note         = {Accessed: 2026-04-02}
}

@misc{nvidia_cutile,
  author       = {{NVIDIA Corporation}},
  title        = {cuTILE Python API Documentation – Execution and JIT Compilation},
  year         = {2026},
  howpublished = {\url{https://docs.nvidia.com/cuda/cutile-python/execution.html}},
  note         = {Accessed: 2026-04-02}
}

@misc{zhou2026,
      title={GPU Accelerated Minimal Auxiliary Basis Approach TDDFT for Large Organic Molecules}, 
      author={Zehao Zhou and Xiaojie Wu and Yanheng Li and Xinran Wei and Cheng Fan and Fusong Ju and Qiming Sun and Yi Qin Gao},
      year={2026},
      eprint={2603.29257},
      archivePrefix={arXiv},
      primaryClass={physics.chem-ph},
      url={https://arxiv.org/abs/2603.29257}, 
}

\end{document}